\documentclass[english,aps,reprint,superscriptaddress]{revtex4-1}
\usepackage[T1]{fontenc}
\usepackage[utf8]{inputenc}
\usepackage{xcolor}
\usepackage{pdfcolmk}
\usepackage{babel}
\usepackage{refstyle}
\usepackage{amsmath}
\usepackage{amssymb}
\usepackage{graphicx}
\usepackage{esint}
\PassOptionsToPackage{normalem}{ulem}
\usepackage{ulem}
\usepackage[unicode=true,pdfusetitle,
 bookmarks=true,bookmarksnumbered=false,bookmarksopen=false,
 breaklinks=false,pdfborder={0 0 1},backref=false,colorlinks=false]
 {hyperref}

\makeatletter

\AtBeginDocument{\providecommand\figref[1]{\ref{fig:#1}}}
\AtBeginDocument{\providecommand\eqref[1]{\ref{eq:#1}}}
\AtBeginDocument{\providecommand\secref[1]{\ref{sec:#1}}}
\AtBeginDocument{\providecommand\subref[1]{\ref{sub:#1}}}
\providecommand{\tabularnewline}{\\}
\RS@ifundefined{subref}
  {\def\RSsubtxt{section~}\newref{sub}{name = \RSsubtxt}}
  {}
\RS@ifundefined{thmref}
  {\def\RSthmtxt{theorem~}\newref{thm}{name = \RSthmtxt}}
  {}
\RS@ifundefined{lemref}
  {\def\RSlemtxt{lemma~}\newref{lem}{name = \RSlemtxt}}
  {}

\providecolor{lyxadded}{rgb}{0,0,1}
\providecolor{lyxdeleted}{rgb}{1,0,0}

\usepackage{lmodern}  
\usepackage{xcolor}
\usepackage{soul}
\usepackage{siunitx}

\providecolor{lyxadded}{rgb}{0,0,1}
\providecolor{lyxdeleted}{rgb}{1,0,0}

\makeatother

\begin{document}

\title{Quantum statistics of polariton parametric interactions}

\date{\today}

\author{M. Sassermann}

\affiliation{Institut für Experimentalphysik, Universität Innsbruck, Technikerstraße
25, A-6020 Innsbruck, Austria}

\author{Z. Vörös}

\affiliation{Institut für Experimentalphysik, Universität Innsbruck, Technikerstraße
25, A-6020 Innsbruck, Austria}

\author{M. Razavi}

\affiliation{School of Electronic and Electrical Engineering, University of Leeds,
Leeds, LS2 9JT, United Kingdom}

\author{W. Langbein}

\affiliation{School of Physics and Astronomy, Cardiff University, Cardiff, CF24
3AA, United Kingdom}

\author{G. Weihs}

\affiliation{Institut für Experimentalphysik, Universität Innsbruck, Technikerstraße
25, A-6020 Innsbruck, Austria}
\begin{abstract}
Using a high-quality GaAs planar microcavity, we optically generate
polariton pairs, and verify their correlations by means of time-resolved
single-photon detection. We find that correlations between the different
modes are consistently lower than identical mode correlations, which
is attributed to the presence of uncorrelated background. We discuss
a model to quantify the effects of such a background on the observed
correlations. Using spectral and temporal filtering, the background
can be suppressed and a change in photon statistics towards non-classical
correlations is observed. These results improve our understanding
of the statistics of polariton-polariton scattering and background
mechanisms, and pave the way to the generation of entangled polariton
pairs.

\end{abstract}
\maketitle

\section{Introduction}

While photons are the most obvious carriers for long-distance communication,
using them in a quantum optical context is challenging. In order to
realize strongly correlated photon pairs, two-qubit gates, quantum
non-demolition measurements, and other quantum protocols, strong interactions
between particles are required, but for photons interactions are extremely
weak in vacuum \cite{Heisenberg1936}. However, through mixing photons
with interacting matter quantum particles one can obtain both strong
interactions and long-distance information transport by tuning the
interaction strength at will \cite{Roy2017}. Several theoretical
works have shown that there are a multitude of schemes to create entangled
states using microcavity polaritons, which are coherent superpositions
of photons and quantum well (QW) excitons, ranging from polarization
\cite{Einkemmer2015a} or energy-time entanglement \cite{Ciuti2004}
to hyper-entanglement \cite{Portolan2013} and cluster states \cite{Pagel2013a,Liew2011}
raising interest in the quantum optics community. On the one hand,
the photonic polariton component provides a direct interface to conventional
optical methods for creation, information encoding, and detection
of quantum states. On the other hand, their excitonic component provides
strong interactions, much larger than in conventional non-linear optical
crystals, which generate correlations. As the main interface to this
system, external photons are coherently converted into polaritons
in the microcavity and vice versa, conserving momentum, polarization
and energy. Because of their interactions, polaritons can scatter
with each other, changing their momentum, polarization or energy states.
The polaritons are converted back into photons with a rate given by
the photon fraction in the polariton divided by the microcavity emission
lifetime. Because of this property one can regard polaritons as strongly
interacting photons \cite{Carusotto2013}. 

In fact, according to recent theoretical predictions, in properly
designed samples, the interaction can be sizable (larger than the
polariton linewidth) even at the single-photon level, which has interesting
implications in quantum optics \cite{Verger2006}. However, until
very recently \cite{Delteil2018,Cuevas2018,Munoz-Matutano2017}, there
has been no unambiguous demonstration of a single-polariton level
non-linearity, and the values that can be reached in these confined
systems are still to be explored.
\begin{figure}[h]
\includegraphics[width=0.98\columnwidth]{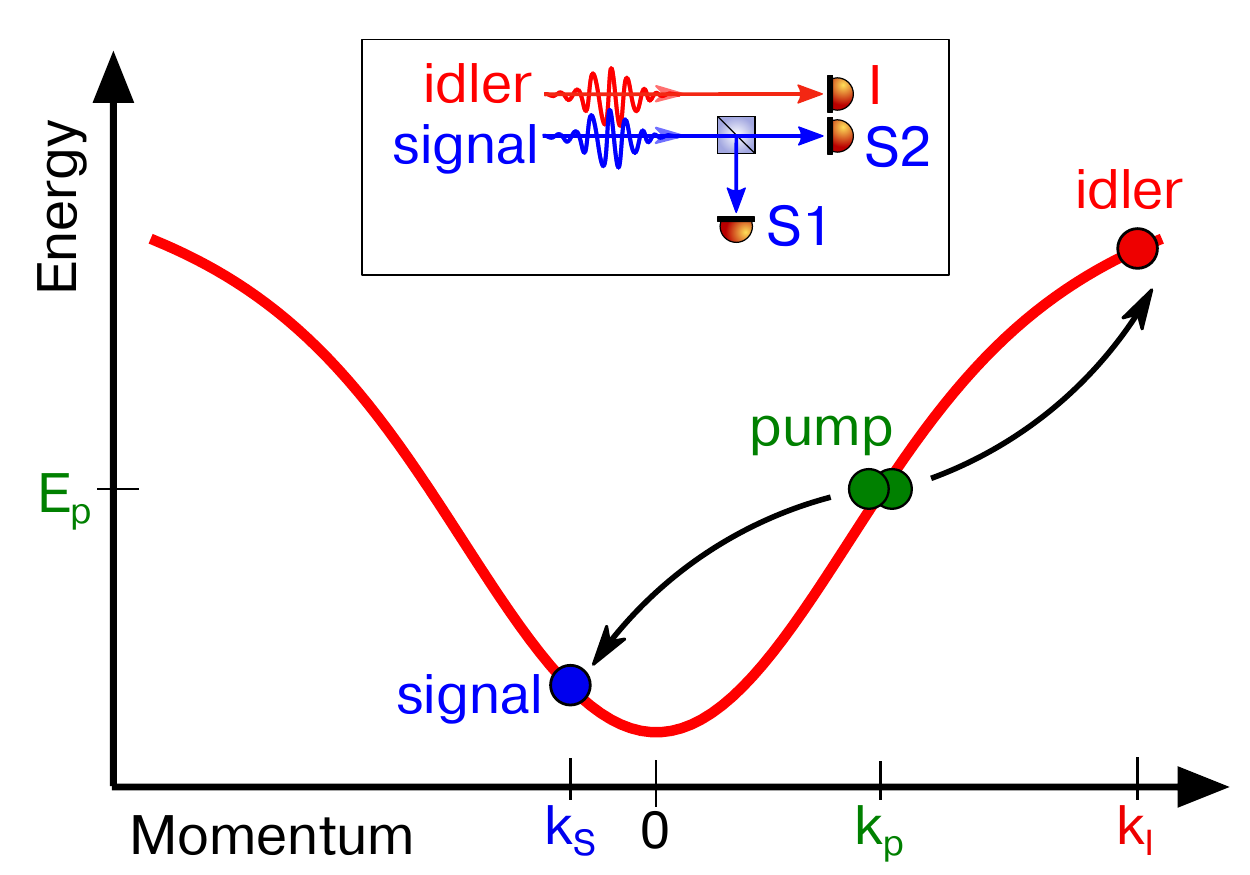}

\caption{Illustration of the parametric scattering process $\left\{ \mathbf{k}_{\mathrm{p}},\mathbf{k}_{\mathrm{p}}\right\} \rightarrow\left\{ \mathbf{k}_{\mathrm{\mathrm{S}}},\mathbf{k}_{\mathrm{I}}\right\} $
on the lower polariton branch, where two pump polaritons (green) scatter
into one signal (red) and one idler polariton (blue). The inset shows
how the modes are split for a Hanbury-Brown and Twiss measurement
on the signal mode.}

\label{fig:param_scatt_schem}
\end{figure}

Polariton scattering can be understood as a four-wave mixing process
in non-linear optics: two pump polaritons are annihilated, and the
signal/idler polaritons are created. The excited polaritons convert
into photons on a picosecond timescale. The degenerate pump case,
when using two identical pump polaritons, as shown in \figref{param_scatt_schem},
is analog to the spontaneous parametric down-conversion process (SPDC).
In the high excitation regime, the pump has an average polariton number
much larger than one and can thus be treated as classical field $P$,
with an interaction Hamiltonian of polariton-polariton scattering
written as \cite{Ciuti2003} 
\begin{equation}
H_{\mathrm{\mathrm{int}}}^{\mathrm{pp}}=Up_{\mathrm{s}}^{\dagger}p_{\mathrm{i}}^{\dagger}p_{\mathrm{p}}^{\phantom{\dagger}}p_{\mathrm{p}}^{\phantom{\dagger}}\approx Up_{\mathrm{s}}^{\dagger}p_{\mathrm{i}}^{\dagger}P^{2}\,.\label{eq:four-wave-mixing-Hamiltonian}
\end{equation}
Here, the operators $p_{\mathrm{p,s,i}}$ destroy a pump, signal,
and idler polariton, respectively and $U$ is the nonlinear interaction
strength. The structure of $H_{\mathrm{\mathrm{int}}}^{\mathrm{pp}}$
is similar to that of the Hamiltonian of a three-wave mixing, or spontaneous
parametric down-conversion process given by \cite{McNeil1983}
\begin{equation}
H_{\mathrm{\mathrm{int}}}^{\mathrm{SPDC}}\propto a_{\mathrm{s}}^{\dagger}a_{\mathrm{i}}^{\dagger}a_{\mathrm{p}}^{\phantom{\dagger}}\approx a_{\mathrm{s}}^{\dagger}a_{\mathrm{i}}^{\dagger}P.\label{eq:spdc-Hamiltonian}
\end{equation}
The only difference is in the coefficient of the pair operator $a_{\mathrm{s}}^{\dagger}a_{\mathrm{i}}^{\dagger}$,
which is, for SPDC, proportional to the driving field, while it is
proportional to the square of the field for polariton scattering.
Similar conclusions can also be drawn in the regime of a weak pump.
Therefore, it is instructive to compare correlation functions obtained
in the case of polaritons to those predicted and measured for SPDC
processes \cite{Giacobino2005}. 

Several experimental studies have found signatures of quantum behavior
in polariton scattering. In continuous excitation settings, Karr et
al. \cite{Karr2004a,Karr2005}, Romanelli et al. \cite{Romanelli2005,Romanelli2010},
and Boulier et al. \cite{Boulier2014} demonstrated squeezing at the
level of a few per cent in the intensity fluctuations of the signal
and idler polaritons (measured in the emitted photons). However, given
the low level of squeezing, the quantumness produced in these experiments
is not enough for any practical applications. Next, Langbein et al.
\cite{Langbein2005,Savasta2005} showed that when certain polariton
scattering pathways are being superposed they result in photon interference.
Finally, Ardizzone et al. \cite{Ardizzone2012a,Ardizzone2013a} measured
second order correlations of photons emitted from polaritonic wires.
These one-dimensional structures have the advantage that the phase
space is reduced, therefore, it is easier to identify paired polaritons
and polariton-polariton interactions are enhanced. Very recently non-classical
correlations were observed \cite{Delteil2018} in 0D structures, which
seem to be a promising platform for studying quantum effects of polaritons.
On the other hand, this very reduction of the phase space leads to
much restricted phase-matching conditions and thus to much larger
exciton and photon linewidths. We therefore choose to work with a
2D microcavity. Unfortunately the interactions between polaritons
(\eqref{four-wave-mixing-Hamiltonian}) were found to be weak with
an interaction strength $U$ on the order of $\approx$\SI{10}{\micro eV \micro m^2}
\cite{Delteil2018,Verger2007} and other effects like luminescence
from bound exciton states \cite{Langbein2005} can be significant.
Understanding the influence of the background processes is crucial
for finding optimal settings for producing maximally correlated photon
pairs.

In this work, we study parametric scattering in planar microcavities
with the aim of producing entangled, or correlated photon pairs. We
measure the quantum statistical properties by means of time-resolved
photon correlation measurements. By employing a Gaussian state model
we can quantify the role of background processes and find optimal
conditions for maximum correlations. Beyond that, we analyze the quantumness
of our source by utilizing non-classicality witnesses. In our measurements
we find a clear link between background processes and the observed
statistics on the boundary between quantum and classical physics.

The article is organized as follows. We begin by introducing Gaussian
theory to compute correlation functions and witnesses in section \ref{sec:theory}.
Before applying the outlined theory to the experimental results in
section \ref{sec:results}, we discuss the experimental setup and
some characteristics of the polariton source, in section \ref{sec:Experimental-details}
and we conclude the paper by identifying paths to increasing the quantumness
of the emission in section \ref{sec:conclusions}.

\section{Gaussian theory of quantum coherence\label{sec:theory}}

In order to be able to produce entangled states from any pair process,
it is necessary for the pairs to exhibit quantum correlations. We
can measure these correlations using photon counting in a Hanbury
Brown-Twiss (HBT) arrangement \cite{BROWN1956} (see \figref{param_scatt_schem}).
To make quantitative predictions about the magnitude of the measured
correlations we employ a model that can be used to compute the underlying
correlation functions and include the effect of uncorrelated background
light superposed with the light from the polariton source.
\begin{figure}[h]
\centering%
\begin{minipage}[t]{0.99\columnwidth}%
\begin{center}
\includegraphics[width=0.98\columnwidth]{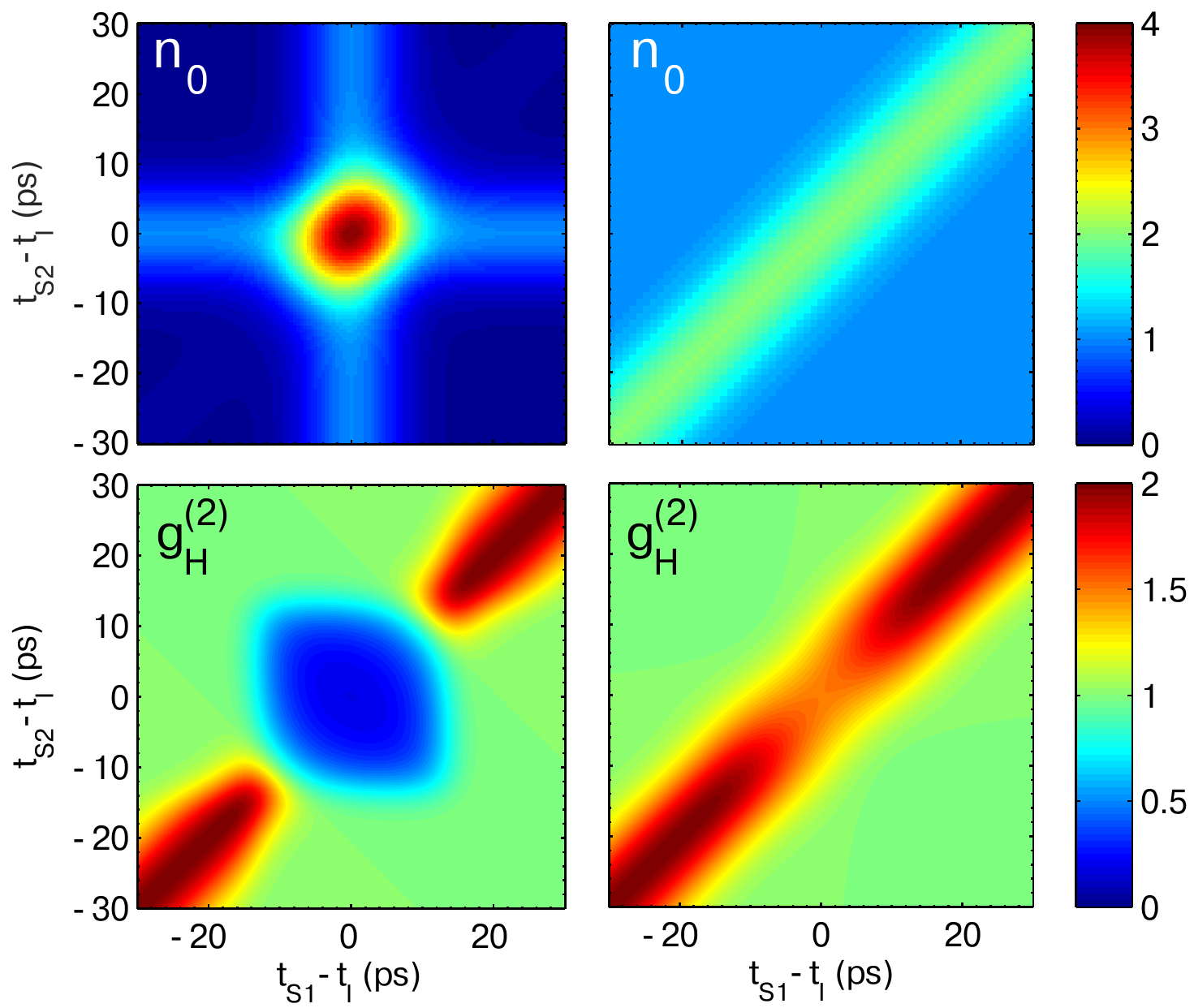}
\par\end{center}%
\end{minipage}\caption{Normalized triple coincidence rates $n_{\text{0}}=N_{\mathrm{SSI}}^{(3)}(t-t_{\mathrm{S1}},t-t_{\mathrm{S\mathrm{2}}},t)/N_{\mathrm{SSI}}^{(3)}(t,\infty,t)$
(top row) and heralded coherence $g_{H}^{^{(2)}}(t-t_{\mathrm{S1}},t-t_{\mathrm{S\mathrm{2}}},t)$
(bottom row) using the quantum correlators from \eqref{covariances}
(left column) and the classical correlators (right column), with $A_{\mathrm{P}}=0.003$
and $T_{\mathrm{c}}=10\,$ps. }
\label{fig:triplespar}
\end{figure}

We start by assuming that the underlying quantum state produced by
our interaction process is a zero-mean Gaussian bi-photon state, which
is fully characterized by only two non-zero temporal correlations,
given by \cite{Shapiro:94}:
\begin{multline}
N_{\mathrm{SS}}(t_{\mathrm{1}},t_{\mathrm{2}})=N_{\mathrm{II}}(t_{\mathrm{1}},t_{\mathrm{2}})=A_{\mathrm{p}}\exp\left(-\frac{(t_{\mathrm{1}}-t_{\mathrm{2}})^{2}}{2T_{\mathrm{c}}^{2}}\right),\\
N_{\mathrm{SI}}(t_{\mathrm{1}},t_{\mathrm{2}})=\sqrt{\frac{A_{\mathrm{p}}}{T_{\mathrm{c}}}\sqrt{\frac{2}{\pi}}}e^{i\text{\ensuremath{\omega_{\mathrm{p}}\frac{t_{\mathrm{1}}}{2}}}}\exp\left(-\frac{(t_{\mathrm{1}}-t_{\mathrm{2}})^{2}}{T_{\mathrm{\mathrm{c}}}^{2}}\right),\label{eq:covariances}
\end{multline}
where $N_{\mathrm{SS}}$ and $N_{\mathrm{II}}$ are the auto-correlation
in the signal and idler mode fields and $N_{\mathrm{SI}}$ is the
cross-correlation between the two fields at a certain pump amplitude
$A_{\mathrm{p}}$, pump frequency $\omega_{\mathrm{p}}$ and coherence
time $T_{\mathrm{\mathrm{c}}}$. Using Wick's theorem, the correlations
above can be used to construct higher-order coherences $N_{\mathrm{AB}}^{(n)}$,
where $\mathrm{A}$ and $\mathrm{B}$ are arbitrary combinations of
modes above. Since we perform photon counting measurements we are
particularly interested in intensity correlations, which are second-order
correlations ($n=2$) and in particular in the degree of second-order
coherence for all combinations of modes in the HBT setup \cite{PhysRevA.79.035801,0953-4075-42-11-114013}.
For example, in the case of the cross-correlations, the coincidence
rate between the signal and idler modes is given by 
\begin{equation}
N_{\mathrm{SI}}^{(2)}(t_{1},t_{2})=\left|N_{\mathrm{SI}}(t_{1},t_{2})\right|^{2}+N_{\mathrm{SS}}(t_{1},t_{2})N_{\mathrm{II}}(t_{1},t_{2}).\label{eq:doublesdef}
\end{equation}
The second-order coherence is then given by
\begin{align}
g_{\mathrm{SI}}^{^{(2)}}(t_{1},t_{2}) & =\frac{N_{\mathrm{\mathrm{SI}}}^{(2)}(t_{1},t_{2})}{N_{\mathrm{SS}}(t_{1},t_{1})N_{\mathrm{II}}(t_{2},t_{2})}\label{eq:gsidef}\\
 & =1+\frac{\left|N_{\mathrm{SI}}^{(2)}(t_{1},t_{2})\right|^{2}}{N_{\mathrm{SS}}(t_{1},t_{1})N_{\mathrm{II}}(t_{2},t_{2})},\nonumber 
\end{align}
 and the two second-order self-coherences $g_{\mathrm{SS}}^{^{(2)}}$,
$g_{\mathrm{II}}^{^{(2)}}$ are computed analogously. For classical
states the three coherence functions are bounded by a generalization
of the Cauchy-Schwartz inequality, which can be derived from first
principles (see for example \cite{walls2008quantum} for further details)
and is given by:
\begin{equation}
g_{\mathrm{SS}}^{(2)}g_{\mathrm{II}}^{(2)}\geq g_{\mathrm{S1I}}^{(2)}g_{\mathrm{S2I}}^{(2)},\label{eq:cauchy}
\end{equation}
where $g_{\mathrm{S1I}}^{(2)}$ and $g_{\mathrm{S2I}}^{(2)}$ are
the second-order coherence functions obtainable when splitting up
the signal mode by a 50:50 beam splitter, as shown in \figref{param_scatt_schem}.
A violation of this inequality is only possible for quantum states
of light \cite{Kuzmich2003}.

The non-classicality of a weakly pumped SPDC-type source can be accessed
via the heralded second-order coherence, where the signal mode is
split by a 50:50-beam splitter with both arms detected by separate
detectors. In this case the Hanbury-Brown Twiss (HBT) measurement
is conditioned on the occurrence of an idler event and the corresponding
signal photon has two pathways to reach a detector. In the ideal case
this leads to sub-Poissonian anti-bunching \cite{Hong1986}. Using
the correlations given in \eqref{covariances}, we can derive first
the triple-coincidence rate and then the heralded coherence: \begin{widetext}

\begin{eqnarray}
N_{\mathrm{SSI}}^{(3)}(t_{\mathrm{S1}},t_{\mathrm{S2}},t_{\mathrm{I}}) & = & \left[2\mathrm{Re}\left\{ N_{\mathrm{SS}}(t_{\mathrm{S1}},t_{\mathrm{S2}})N_{\mathrm{SI}}(t_{\mathrm{I}},t_{\mathrm{S2}})^{*}N_{\mathrm{SI}}(t_{\mathrm{S1}},t_{\mathrm{I}})\right\} \right.+\nonumber \\
 &  & +N_{\mathrm{SS}}\mathrm{(t_{\mathrm{S1}},\mathrm{t}_{S1})}\left|N_{\mathrm{SI}}(t_{\mathrm{S2}},t_{\mathrm{I}})\right|^{2}+N_{\mathrm{SS}}(t_{\mathrm{S2}},t_{\mathrm{S2}})\left|N_{\mathrm{SI}}(t_{\mathrm{S1}},t_{\mathrm{I}})\right|^{2}\nonumber \\
 &  & +\left.N_{\mathrm{II}}(t_{\mathrm{I}},t_{\mathrm{I}})\left(\left|N_{\mathrm{SS}}(t_{\mathrm{S1}},t_{\mathrm{S2}})\right|^{2}+N_{\mathrm{SS}}(t_{\mathrm{S1}},t_{\mathrm{S1}})N_{\mathrm{SS}}(t_{\mathrm{S2}},t_{\mathrm{S2}})\right)\right]N_{\mathrm{II}}(t_{\mathrm{I}},t_{\mathrm{I}}),\label{eq:triplesdef}\\
g_{\mathrm{H}}^{^{(2)}}(t_{\mathrm{S1}},t_{\mathrm{S2}},t_{\mathrm{I}}) & \equiv & \frac{\left\langle \hat{a}_{\mathrm{I}}^{\dagger}(t_{\mathrm{I}})\hat{a}_{\mathrm{S1}}^{\dagger}(t_{\mathrm{S1}})\hat{a}_{S2}^{\dagger}(t_{\mathrm{S2}})\hat{a}_{\mathrm{S2}}(t_{\mathrm{S2}})\hat{a}_{S1}(t_{\mathrm{S1}})\hat{a}_{\mathrm{I}}(t_{\mathrm{I}})\right\rangle \left\langle \hat{a}_{\mathrm{I}}^{\dagger}(t_{\mathrm{I}})\hat{a}_{\mathrm{I}}(t_{\mathrm{I}})\right\rangle }{\left\langle \hat{a}_{\mathrm{I}}^{\dagger}(t_{\mathrm{I}})\hat{a}_{\mathrm{S1}}^{\dagger}(t_{\mathrm{S1}})\hat{a}_{S1}(t_{\mathrm{S1}})\hat{a}_{\mathrm{I}}(t_{\mathrm{I}})\right\rangle \left\langle \hat{a}_{\mathrm{I}}^{\dagger}(t_{\mathrm{I}})\hat{a}_{\mathrm{S2}}^{\dagger}(t_{\mathrm{S2}})\hat{a}_{S2}(t_{\mathrm{S2}})\hat{a}_{\mathrm{I}}(t_{\mathrm{I}})\right\rangle }=\frac{N_{\mathrm{SSI}}^{(3)}(t_{\mathrm{S1}},t_{\mathrm{S2}},t_{\mathrm{I}})}{N_{\mathrm{SI}}^{(2)}(t_{\mathrm{S1}},t_{\mathrm{I}})N_{\mathrm{SI}}^{(2)}(t_{\mathrm{S2}},t_{\mathrm{I}})},\label{eq:g2hdef}
\end{eqnarray}
 \end{widetext}where $N^{*}$ denotes the complex conjugate of the
correlations and $g_{\mathrm{H}}^{^{(2)}}$ conditions the the second-order
self-coherence on the measurement of an idler photon 
\begin{figure*}[t]
\centering%
\begin{minipage}[t]{0.98\textwidth}%
\begin{center}
\includegraphics[width=0.98\textwidth]{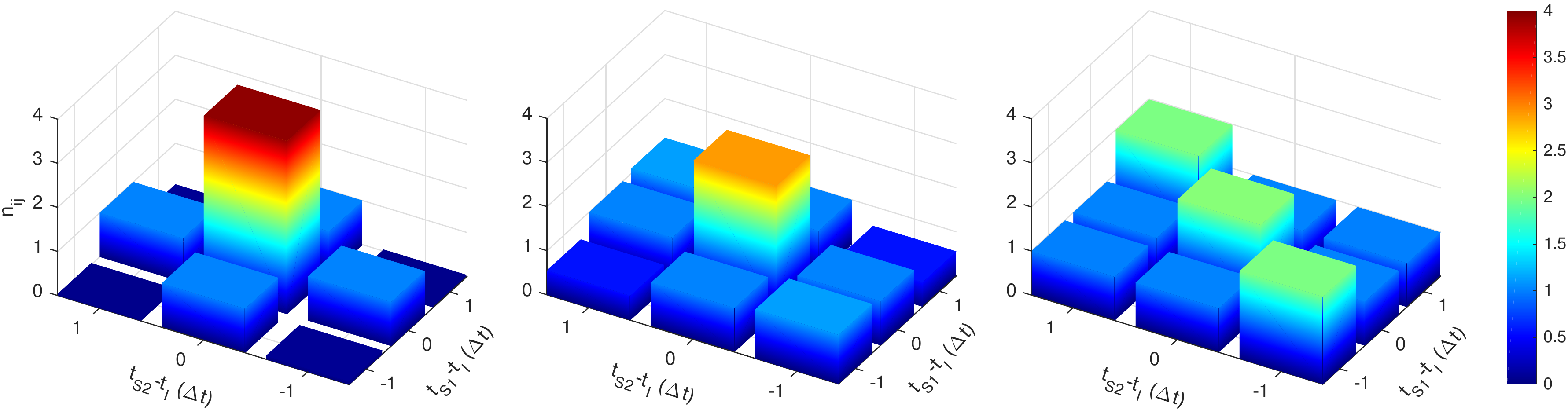}
\par\end{center}%
\end{minipage}\caption{Normalized asymptotic value $n_{ij}$ of the normalized triples with
discrete delays $T_{\mathrm{rep}}$ along different axes in time,
computed with our model,1 assuming $T_{\mathrm{rep}}\gg T_{\mathrm{c}}$,
for three different pump amplitudes $A_{\mathrm{p}}=$\SI{0.003}{s^{-1}},
\SI{0.1}{s^{-1}}, \SI{4.7}{s^{-1}}. }
\label{fig:hsitvspumpparam}
\end{figure*}
If the system were in a pure single biphoton state, $N_{\mathrm{SSI}}^{(3)}$
would be zero. One property common to all quantities defined above
is that they only depend on time differences and the most interesting
effects occur when all arrival times are close together.

The arrival-time dependence of the triples and the corresponding $g_{\mathrm{H}}^{^{(2)}}$
for both the low pump power quantum regime and the high pump power
classical regime is given in \figref{triplespar}. While the theory
above assumes a continuous-wave pump, our measurements were performed
in the pulsed regime, where the basic structure above is still valid,
but only accessible in discrete form (when integrating over the detected
pulses using finite integration windows). In our measurements the
pulse repetition period $T_{\mathrm{rep}}$ is much larger than the
lifetimes of the states ($T_{\mathrm{rep}}\gg T_{\mathrm{c}}$). Thus
we can still use the results from above to characterize the behavior
at zero time delay and asymptotically from the temporal distribution
shown in \figref{triplespar} at large time delays in any direction.
In order to capture the change from classical to quantum correlations
in this discrete framework we introduce the ratio $n_{0}$ between
the value of the central peak $N_{\mathrm{SSI}}(0,0,0)$ normalized
to the asymptotic value $N_{\mathrm{SSI}}(0,\infty,0)$. For purely
parametric light this ratio approaches different constant values,
depending on the pump regime, as given by\begin{widetext} 
\begin{equation}
n_{0}=\frac{N_{\mathrm{SSI}}^{(3)}(0,0,0)}{N_{\mathrm{SSI}}^{(3)}(0,T_{\mathrm{rep}},0)}\overset{T_{\mathrm{rep}}\gg T_{\mathrm{c}}}{\approx}\frac{4\sqrt{2/\pi}+2A_{\mathrm{p}}T_{\mathrm{c}}}{\sqrt{2/\pi}+A_{\mathrm{p}}T_{\mathrm{c}}}\approx\begin{cases}
4, & A_{\mathrm{p}}\ll1\\
2, & A_{\mathrm{p}}\gg1
\end{cases}.\label{eq:triprat0quant}
\end{equation}
\end{widetext}For large pump amplitudes ($A_{\mathrm{p}}\gg1$) the
value of $n_{0}$ is identical to the maximum value obtainable for
a classical single-mode thermal state split into a ``signal'' and
an ``idler'' mode. In the case of the classical parametric amplifier
introduced in \cite{Shapiro:94} the fields are composed of zero-mean
jointly Gaussian random processes, where signal and idler photons
are drawn from a bivariate Gaussian distribution. For this case we
obtain the maximum value given by:
\begin{equation}
n_{0}^{\mathrm{class}}=\frac{6A_{\mathrm{p}}^{3}}{2A_{\mathrm{p}}^{3}(1+2\exp{-\frac{\text{\ensuremath{T_{\mathrm{rep}}^{2}}}}{T_{\mathrm{c}}}})}\overset{T_{\mathrm{rep}}\gg T_{\mathrm{c}}}{\approx}3,\label{eq:triprat0class}
\end{equation}
independent of the pump amplitude. 

\begin{figure}[h]
\centering%
\begin{minipage}[t]{0.95\columnwidth}%
\begin{center}
\includegraphics[width=0.98\columnwidth]{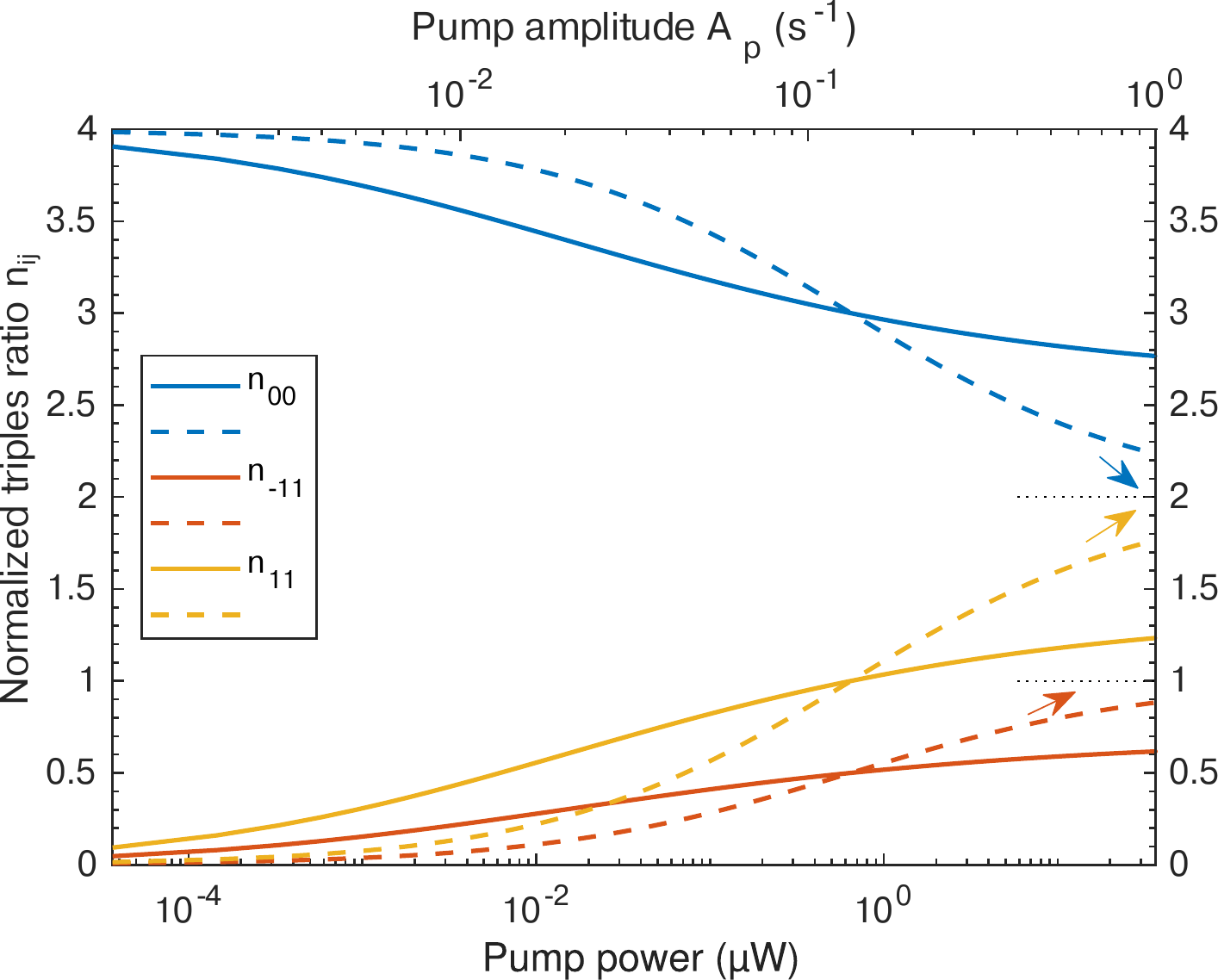}
\par\end{center}%
\end{minipage}\caption{Theoretical pump power dependence of $n_{ij}$, as labeled, without
background $\alpha=1$ (dashed lines) and with $\alpha=1/3$ (solid
lines), having two thirds of the total counts from thermal background.
The arrows indicate the asymptotic values towards the curves are converging
towards.}
\label{fig:mixedlightdependencies}
\end{figure}

In the pulsed case, the relevant temporal structure of the correlations
can be captured with the time arguments discretized to multiples of
the pulse repetition period,

\[
n_{ij}\equiv\frac{N_{\mathrm{SSI}}(iT_{\mathrm{rep}},jT_{\mathrm{rep}},0)}{N_{\mathrm{SSI}}(0,T_{\mathrm{rep}},0)},
\]
for $i,j\in\left\{ -1,0,1\right\} $ and $T_{\mathrm{rep}}\gg T_{\mathrm{c}}$.
The resulting $n_{ij}$ is shown in in \figref{hsitvspumpparam} for
different pump amplitudes. For high pump powers the normalized triples
converge to the values for purely classical thermal light as is detailed
in \figref{mixedlightdependencies}.

In the framework established so far, a value $n_{0}>3$ would indicate
non-classical behavior, however, care must be taken because the underlying
theory is not general, since we assume particular statistical processes.
In order to quantify the quantumness of our source in a general way
we are going to analyze the measured photon emission probabilities,
as outlined in the following section. In a real experimental setting
the fluorescence from other excitations of the system or host material
may cause strong background contributions in both the signal and the
idler mode. Given that we know the nature of the background we can
directly incorporate background light in the above model, as we will
show in the next section.

\subsection{Correlations of mixed light}

In the following discussion we consider the case of a uniform background
emission from a thermal source superimposed on the parametric emission.
The main processes leading to background emission are all incoherent,
because there is at least one scattering partner with random phase
involved (phonons or excitons from the high k-vector reservoir). A
coherent contribution could come from resonant Rayleigh scattering
\cite{Langbein2002}, but we strongly reduce it by spectral, spatial
and polarization filtering (see, \secref{Experimental-details}) and
can therefore neglect it. We therefore consider the dephasing to be
fast enough to act as incoherent background light on the experimentally
relevant time scales. In order to model this behavior we assume that
an operator $\hat{F}_{\mathrm{th}}$ creates additional photons in
the same states as above leading to an effective photon annihilation
operator $\hat{a}_{\mathrm{tot}}=\alpha\hat{a}+\sqrt{1-\alpha^{2}}\hat{F}_{\mathrm{th}}$,
where $\hat{F}_{\mathrm{th}}$ obeys single-mode thermal state statistics
with coherence length $T_{\mathrm{th}}$ as defined by the following
correlators:
\begin{eqnarray*}
N_{\mathrm{th}}(t_{1},t_{2}) & = & A_{\mathrm{th}}e^{-i\text{\ensuremath{\omega_{\mathrm{p}}t_{1}}}}\exp\left(-\frac{(t_{1}-t_{2})^{2}}{2T_{\mathrm{th}}^{2}}\right),\\
N_{\mathrm{SI},\mathrm{th}}(t_{1},t_{2}) & = & 0.
\end{eqnarray*}
The cross-correlations are zero because $\left\langle F_{i}F_{j}\right\rangle _{\mathrm{th}}=\left\langle F_{i}^{\dagger}F_{j}^{\dagger}\right\rangle _{\mathrm{th}}=0$.
If we set all mixed terms like $\left\langle \hat{a}^{\dagger}\hat{F}_{\mathrm{th}}\right\rangle $
to zero, corresponding to independent photon sources, the total two-time
correlators are given by:
\begin{eqnarray}
N_{SS,\mathrm{tot}}(t_{1},t_{2}) & = & N_{SS}(t_{1},t_{2})+N_{SS,\mathrm{th}}(t_{1},t_{2}),\nonumber \\
N_{SI,\mathrm{tot}}(t_{1},t_{2}) & = & N_{SI}(t_{1},t_{2}).\label{eq:bgcorrs_total}
\end{eqnarray}
The contribution of each term depends on the contribution of parametric
emission to the total emission given by the ratio $\alpha=A_{\mathrm{p}}/(A_{\mathrm{th}}+A_{\mathrm{p}})$.
We can now use these correlators in the same way as in \eqref{covariances}
and compute the higher order correlations. The results from these
calculations are similar to what we discussed above, where $\alpha$
has a similar effect on the correlations as an increase of pump intensity
- the higher the background contribution the lower the value of $n_{0}$
becomes until it approaches a value of three. We furthermore find
that as the background becomes dominant, the power dependence of $n_{ij}$
becomes flatter, because of the power-independent cross-correlations
of the thermal state. The conditional coherence for mixed light behaves
in a similar way, where a gradual increase of the background contribution
results in less pronounced anti-bunching and eventually bunching.
The exact crossover depends on the driving strength of the parametric
amplifier. 
\begin{figure}[h]
\centering%
\begin{minipage}[t]{0.95\columnwidth}%
\begin{center}
\includegraphics[width=0.98\columnwidth]{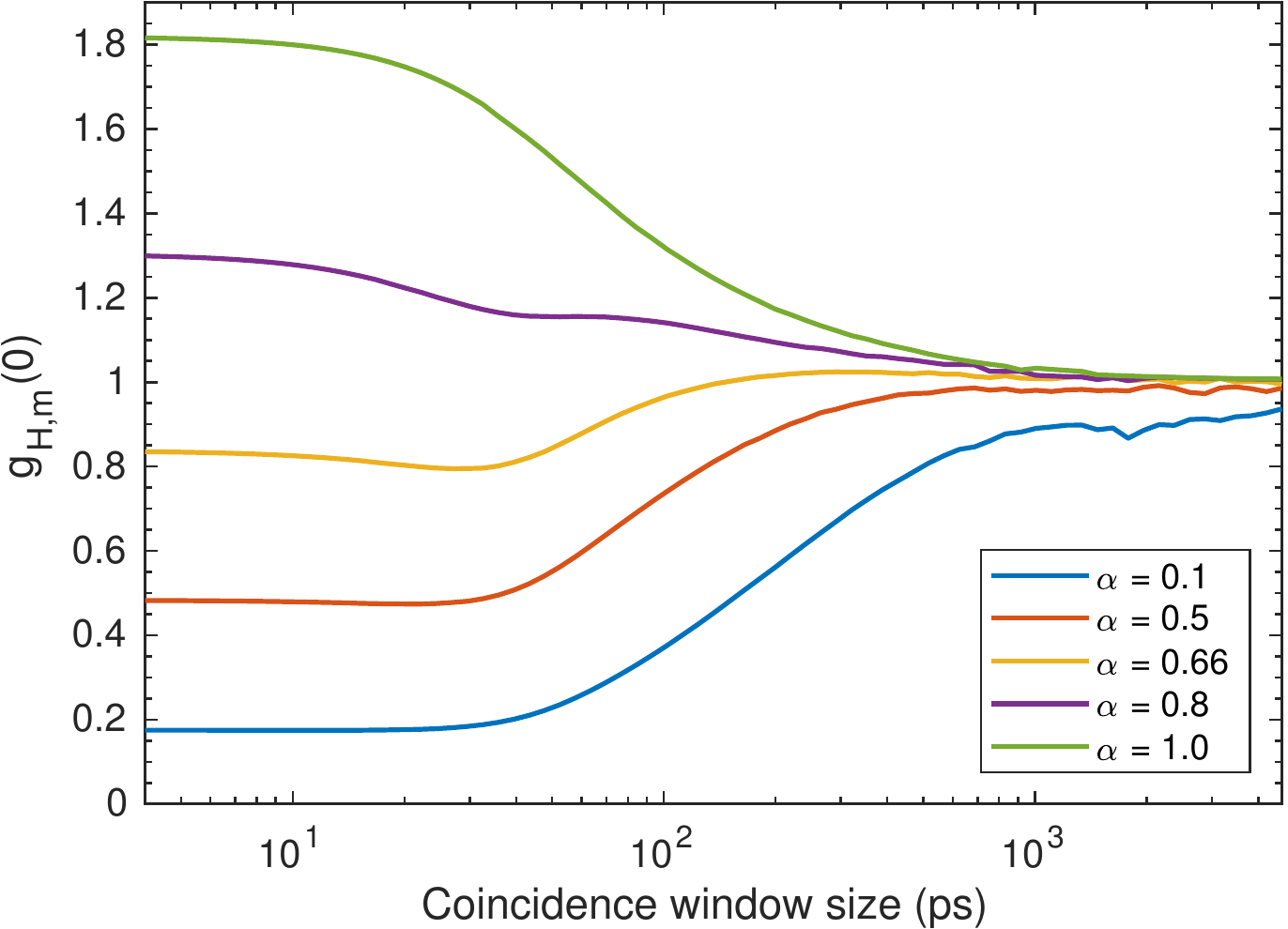}
\par\end{center}%
\end{minipage}\caption{Calculated coincidence window dependence of the conditional coherence
function $g_{\mathrm{H,m}}$ for mixed light with different $\alpha$.
The parameters are: the detector time resolution $T_{\mathrm{S1}}=T_{\mathrm{S2}}=T_{\mathrm{I}}=35\,$ps
(FWHM), $T_{\mathrm{th}}$=$100\,$ps (FWHM), $T_{\mathrm{c}}=10\,$ps
(FWHM), and $A_{\mathrm{p}}=$\SI{0.001}{s^{-1}}. }
\label{fig:ghmcoincdep}
\end{figure}

\subsection{Including the temporal instrument response function}

In order to include the effect of the finite temporal response of
the single photon detectors we convolve all the above quantities with
the instrument response function of our detection system. As shown
in \figref{detresponses}, our best detectors exhibit an almost ideal
Gaussian response we can therefore use convolutions with Gaussians
$G(t,\tau)$ with a temporal width $\tau$. For example, the triple
coincidence rate can be computed by the following expression \cite{0953-4075-42-11-114013}:\begin{widetext}
\[
N_{\mathrm{SSI}}^{\mathrm{IRF}}(t_{1,}t_{2},t_{3})=\int\,\mathrm{d}t_{\mathrm{S_{1}}}\int\,\mathrm{d}t_{\mathrm{\mathrm{S}_{2}}}\int\,\mathrm{d}t_{\mathrm{I}}\,G(t_{\mathrm{S_{1}}}-t_{1},\tau_{\mathrm{S1}})G(t_{\mathrm{S_{2}}}-t_{2},\tau_{\mathrm{S2}})G(t_{\mathrm{I}}-t_{3},\tau_{\mathrm{I}}).
\]

\end{widetext}We can now compute the number of triple coincidence
events within a given coincidence window $\left[t-\tau_{\mathrm{W}},t+\tau_{\mathrm{W}}\right]$
of the measurements by integration over this window:
\[
N_{T}(t_{1,}t_{2})=\frac{1}{4\tau_{\mathrm{W}}^{2}}\int_{-\tau_{\mathrm{W}}}^{\tau_{\mathrm{W}}}\,\mathrm{d}t_{\mathrm{1}}\int_{-\tau_{\mathrm{W}}}^{\tau_{\mathrm{W}}}\,\mathrm{d}t_{\mathrm{2}}\,N_{\mathrm{SSI}}^{\mathrm{IRF}}(t_{1,}t_{2},0),
\]
where we have set $t_{3}=0$ for simplicity. All other quantities
can be computed analogously and treated in the same way as described
above. In order to illustrate the effect of the finite instrument
response and the coincidence window we show in \figref{ghmcoincdep}
the conditional coherence function $g_{\mathrm{H,m}}$ of a mixture
of parametric and thermal light after convolution and integration.
It is evident that a coincidence window much smaller than the detector
time resolution does not have any effect, whereas larger coincidence
windows, greater than the coherence time, result in the signal becoming
increasingly uncorrelated.

\subsection{Nonclassicality characterization using witnesses\label{sub:Non-classicality-characterizatio}}

A state is nonclassical if it cannot be expressed by any statistical
mixture of coherent states \cite{Glauber1963}. A quantity for characterizing
the nonclassicality of single-photon states, as introduced in \cite{PhysRevLett.113.223603},
is defined in terms of photon emission probabilities ($P_{0}=\langle0|\rho|0\rangle$,$P_{1}=\langle1|\rho|1\rangle$
and $P_{2+}=1-P_{0}-P_{1}$) of a system described by the density
matrix $\rho$. Based on this, two witnesses are defined \cite{PhysRevLett.113.223603}:

\begin{eqnarray}
P_{2+}<\frac{1}{2}P_{1}^{2}, &  & P_{2+}<\frac{2}{3}P_{1}^{3},\label{eq:QNC}
\end{eqnarray}
 where the first inequality defines the non-classicality (NC) witness.
If the measured probabilities fulfill the inequality, the underlying
photon state is nonclassical. The second inequality is even more restrictive
and defines the upper bound for the underlying quantum state to be
additionally quantum non-Gaussian. A state is quantum non-Gaussian,
if its Wigner function representation exhibits negativity. This means
that the non-classicality witness can always certify that a given
state is non-classical, but false negatives are possible when $P_{2+}\gtrsim P_{1}$
\cite{PhysRevLett.113.223603}, for example when detecting two-photon
Fock-states. We can compute the probabilities directly from the photon
detection, as we will show later in this section. 
\begin{figure}
\centering%
\begin{minipage}[t]{0.95\columnwidth}%
\includegraphics[width=0.98\columnwidth]{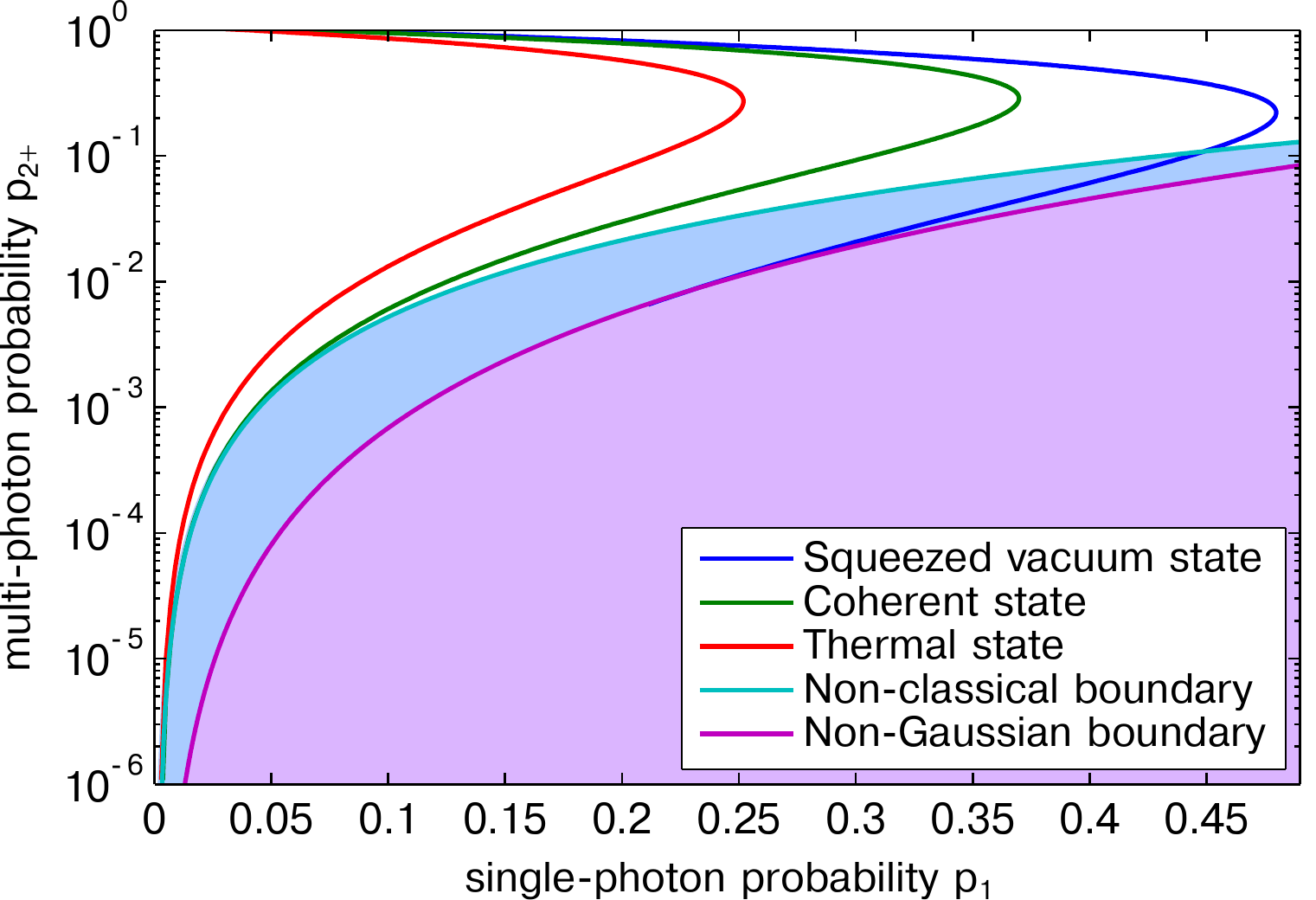}%
\end{minipage}

\caption{Calculated multiple versus single photon probability for different
states, as labeled. The shaded blue area marks the non-classical (NC),
but quantum Gaussian region. The shaded purple area marks the non-classical
region that is quantum non-Gaussian (QNG) at the same time, fulfilling
both inequalities given by \eqref{QNC}. Note that this is a parametric
plot as a function of $\text{\ensuremath{\mu}}$ over the full range
including the cases where $P_{2+}>P_{1}$ and therefore there are
two solutions for each value of $P_{1}$.}

\label{fig:spmptheory}
\end{figure}

For better illustration of the meaning of the above inequalities,
we take three different states: a squeezed state from a parametric
down conversion source, a single-mode thermal state and a coherent
state from an ideal laser. The multiple versus single photon probability
of the different states are shown in \figref{spmptheory}. The probabilities
only depend on the mean photon number $\mu$. In the regime of a weakly
pumped source emitting low photon numbers ($\mu\ll1$), $P_{2+}<\frac{1}{2}P_{1}^{2}$
is satisfied only by the squeezed state, while the coherent state
of the laser marks exactly the boundary between classical and nonclassical
(as given by \eqref{QNC}) and the thermal state is always above this
boundary, see \figref{spmptheory}. The squeezed vacuum state above
converges to the quantum non-Gaussian boundary for very small photon
numbers, but still remains Gaussian. In the case of high average photon
numbers the criterion does not allow the distinction between quantum
and classical states.

The probabilities $P_{1}$ and $P_{2+}$ can be estimated directly
from the measured detector counts in a coincidence-measurement, see
\figref{experimental_setup} and \cite{PhysRevLett.107.213602} for
more details. Important for the derivation of the probabilities is
only the total transmittance $\mathcal{T}$ (including losses and
imbalance and relative detector efficiencies) of the last beam splitter
in \figref{experimental_setup}, splitting up the signal mode. The
estimation of $P_{1}$ depends on the effective imbalance of the two
detection channels $\mathcal{T}/(1-\mathrm{\mathcal{T}})$ and if
we assume, for example, that $\mathcal{T}>1/2$, we can give an upper
bound on the transmittance, where $\mathcal{T\leq T}_{\mathrm{e}}$
from the measured count rates \cite{PhysRevLett.107.213602}:
\[
\mathcal{T}_{\mathrm{e}}=\frac{N_{\mathrm{S1,I}}}{N_{\mathrm{S1,I}}+N_{\mathrm{S2,I}}},
\]
where $N_{\mathrm{S\mathit{n},I}}$ are the double coincidence counts
between the signal channels and the idler channel for $n=1,2$. As
derived in \cite{PhysRevLett.107.213602}, we can estimate $P_{1}$
by the experimentally measurable quantity 
\[
P_{1}^{e}=\frac{N_{\mathrm{S1,I}}+N_{\mathrm{S2,I}}}{N_{\mathrm{I}}}-\frac{N_{\mathrm{S1,S2,I}}(1-\mathcal{T}_{\mathrm{e}})^{2}+\mathcal{T}_{\mathrm{e}}^{2}}{2N_{\mathrm{I}}T_{\mathrm{e}}(1-\mathcal{T}_{\mathrm{e}})},
\]
where $P_{1}^{e}\leq P_{1}$ with an error on the order of $P_{3}$
\cite{PhysRevLett.107.213602} and $N_{\mathrm{S1,S2,I}}$ and $N_{I}$
are the triples and idler-trigger counts, respectively. The probability
to detect the vacuum state is given by:

\[
P_{0}^{e}=1-\frac{N_{\mathrm{S1,I}}+N_{\mathrm{S2,I}}+N_{\mathrm{S1,S2,I}}}{N_{\mathrm{I}}}.
\]
Putting all of the above together we can calculate the probability
to detect two and more photons:
\[
P_{2+}^{e}=\frac{N_{\mathrm{S1,I}}^{2}+N_{\mathrm{S1,S2,I}}\left(2N_{\mathrm{S1,I}}+N_{\mathrm{S2,I}}\right)N_{\mathrm{S2,I}}}{2N_{\mathrm{I}}N_{\mathrm{S1,I}}N_{\mathrm{S2,I}}}.
\]
We will use these expressions in section \ref{sec:results} to characterize
the light emitted from our polariton source.

In order to quantify the convex distance to non-classical states,
we follow Ref. \cite{PhysRevLett.107.213602} in defining a witness
function given by:

\[
W(a,r)\equiv ap_{0}(\mu)+p_{1}(\mu),
\]
where $a$ is a free parameter for optimization and the probabilities
$p_{0,1}(\mu)$ are computed from an ideal coherent state (Poisson
distribution) with mean photon number $\mu$ and we assume that $p_{0}+p_{1}\leq1$
(low excitation regime). The convex distance from the experimentally
determined probabilities is given by: 
\begin{equation}
\Delta W_{f}(a)=aP_{0}^{\mathrm{e}}+P_{1}^{\mathrm{e}}-W(a,\mu_{\mathrm{opt}}),\label{eq:wittmax}
\end{equation}
 where $\mu_{\mathrm{opt}}=1-a$ is the mean photon number maximizing
$W$. In order to obtain an upper bound for the distance we maximize
$\Delta W_{f}$ with respect to $a$, i.e:

\begin{equation}
\Delta W\equiv\textrm{{max}}(\Delta W_{f}(a)).\label{eq:wittdistmax}
\end{equation}
Further details can be found in Ref. \cite{PhysRevLett.107.213602}
and in the discussion of the experimental data in \subref{High-time-resolution}.

\section{Experimental details\label{sec:Experimental-details}}

The investigated sample is an epitaxially grown semiconductor structure
that contains a $\lambda$ cavity with a single 25\,nm wide GaAs
quantum well (QW) located at the anti node of the cavity field. Details
about the sample can be found in \cite{Jensen2000}. The measured
polariton linewidths are on the order of \SI{120}{\micro eV} and
the Rabi splitting is $3.6\,$meV \cite{Borri2000a}. The corresponding
polariton lifetimes are on the order of 10\,ps \cite{Langbein2002}.
In contrast to most GaAs-based polariton samples reported in the literature,
in our case, the unusually wide QW leads to three visible polariton
branches, because the light hole-heavy hole splitting is only 4\,meV,
and thus, both the light hole and the heavy hole exciton are efficiently
coupled to the cavity. However, we always work on the lowest branch,
and the middle and upper polaritons can safely be neglected. 

\begin{figure}[h]
\includegraphics[width=0.98\columnwidth]{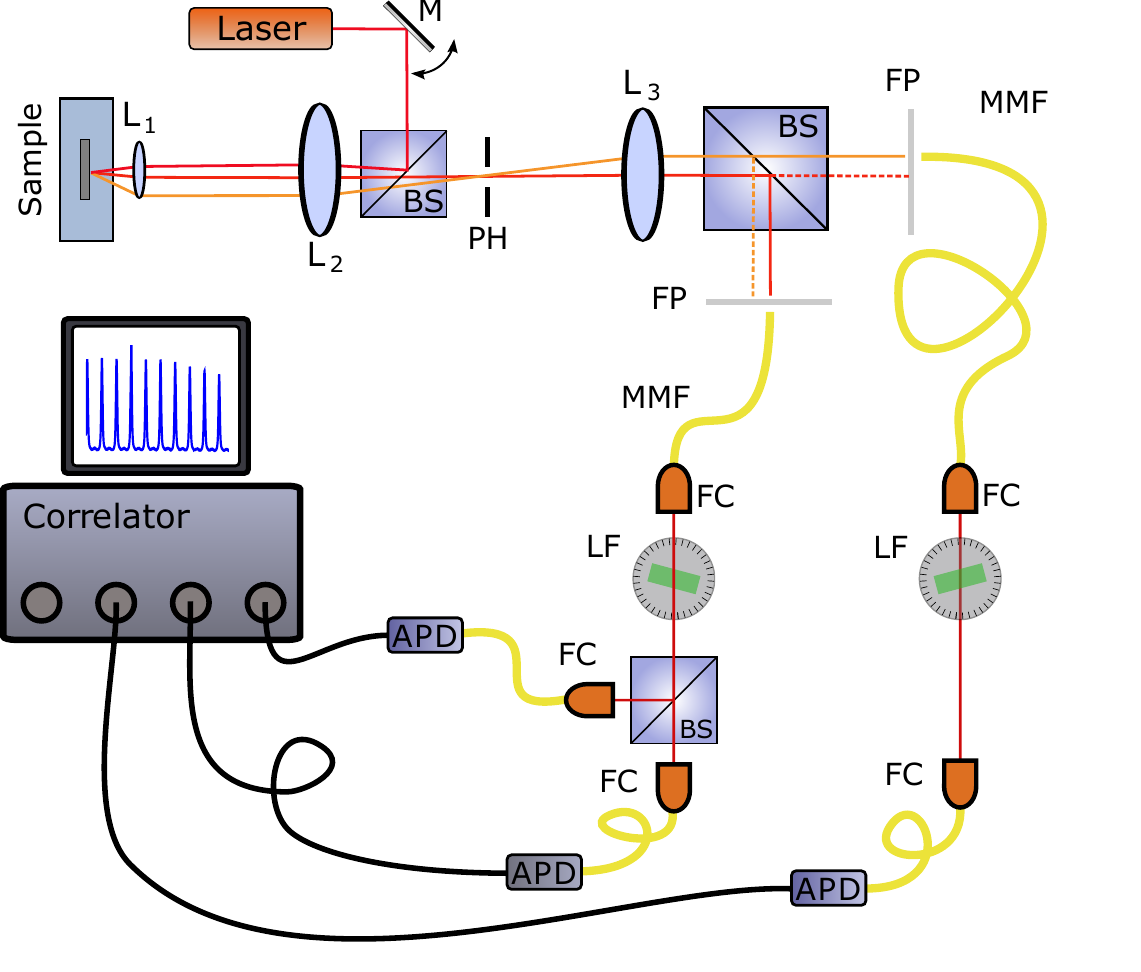}

\caption{Schematic of the experimental setup. Abbreviations: \emph{M:} mirror,
\emph{BS:} non-polarizing beam splitter, $\mathit{\mathcal{\textrm{\ensuremath{L_{1}},\ensuremath{L_{2}},\ensuremath{L_{3}}}}}$:
lenses with focal lengths \SIlist{9;500;500}{mm}, respectively, \emph{PH:}
pinhole, \emph{FP:} Fourier plane, \emph{LF:} line filter, \emph{MMF:}
multi-mode fibre, \emph{FC:} fibre collimator and \emph{APD:} avalanche
photodiode. }

\label{fig:experimental_setup}
\end{figure}

The experimental setup is shown in \figref{experimental_setup}. The
sample is held in a closed-cycle cryostat at temperatures in the range
of 15-20\,K, and is excited resonantly on the lowest branch by short
(3-20\,ps) laser pulses focused onto the sample by an aspheric lens
(Edmund Optics \#67-257), \emph{$\textrm{\ensuremath{L_{1}}}$}, of
focal length $f_{1}=9$ mm. We use a second lens \emph{$\textrm{\ensuremath{L_{2}}}$},
at the distance $f_{1}+f_{2}$ to focus the beam into the back-focal
plane of the first lens. The excitation angle is set by a tiltable
mirror located in the back-focal plane of $\textrm{\ensuremath{L_{2}}}$. 

The luminescence is collected through \emph{$\textrm{\ensuremath{L_{1}}}$},
and passed through an optical relay consisting of two 500-mm focal
length lenses, \emph{$\textrm{\ensuremath{L_{2}}}$}, and \emph{$\textrm{\ensuremath{L_{3}}}$},
respectively. The distance between $\textrm{\ensuremath{L_{2}}}$
and $\textrm{L}_{3}$ is given by $f_{2}+f_{3}$ and produces an image
of the back-focal plane of $\textrm{L}_{1}$ at distance $f_{3}$
from $\textrm{\ensuremath{L_{3}}}$, which are the two Fourier planes
\emph{FP}. In these planes the signal and idler modes are spatially
selected by two multi-mode fibers with \SI{62}{\micro m} core diameter.
The fibers are attached to 2-axis translation stages, and thus, arbitrary
position in the momentum plane, and thus signal and idler modes can
be chosen. 

In order to select the real-space region from which the emission is
collected, we spatially filter the emitted light by a pinhole \textit{PH}
in the common focal plane of the lenses \emph{$\textrm{\ensuremath{L_{2}}}$}
and \emph{$\textrm{\ensuremath{L_{3}}}$}, which is where the real-space
image is formed. Spectral filtering is achieved by means of two interference
line filters, \emph{LF}, of full width at half maximum of 0.25\,nm
and a central frequency around the lower polariton resonance ($\approx817\,$nm).
The filters are mounted on rotation stages to provide angle tuning
in a range of about 6\,nm. Since we turn the filters only by a few
degrees around the center, there are only negligible changes in transmission
line width and changes in polarization are adjusted using fiber polarization
controllers. The light filtered in this way is split once more by
a polarization independent 50:50 beam splitter \textit{BS} in the
case of the signal mode, collected through fibre collimators, matching
the fibers, and then fed into the detectors. The detector signals
are analyzed by a HydraHarp time correlator from PicoQuant. For the
measurements in section \ref{sec:results} we used avalanche photodiodes
with a combined time resolution (detectors and counter electronics)
of roughly $45\,$ps FWHM, while we used superconducting nanowire
detectors with $35\,$ps FWHM for all other measurements. Further
details about the detectors can be found in section \ref{sec:Appendix-B}. 

\begin{figure}[h]
\includegraphics[width=1\columnwidth]{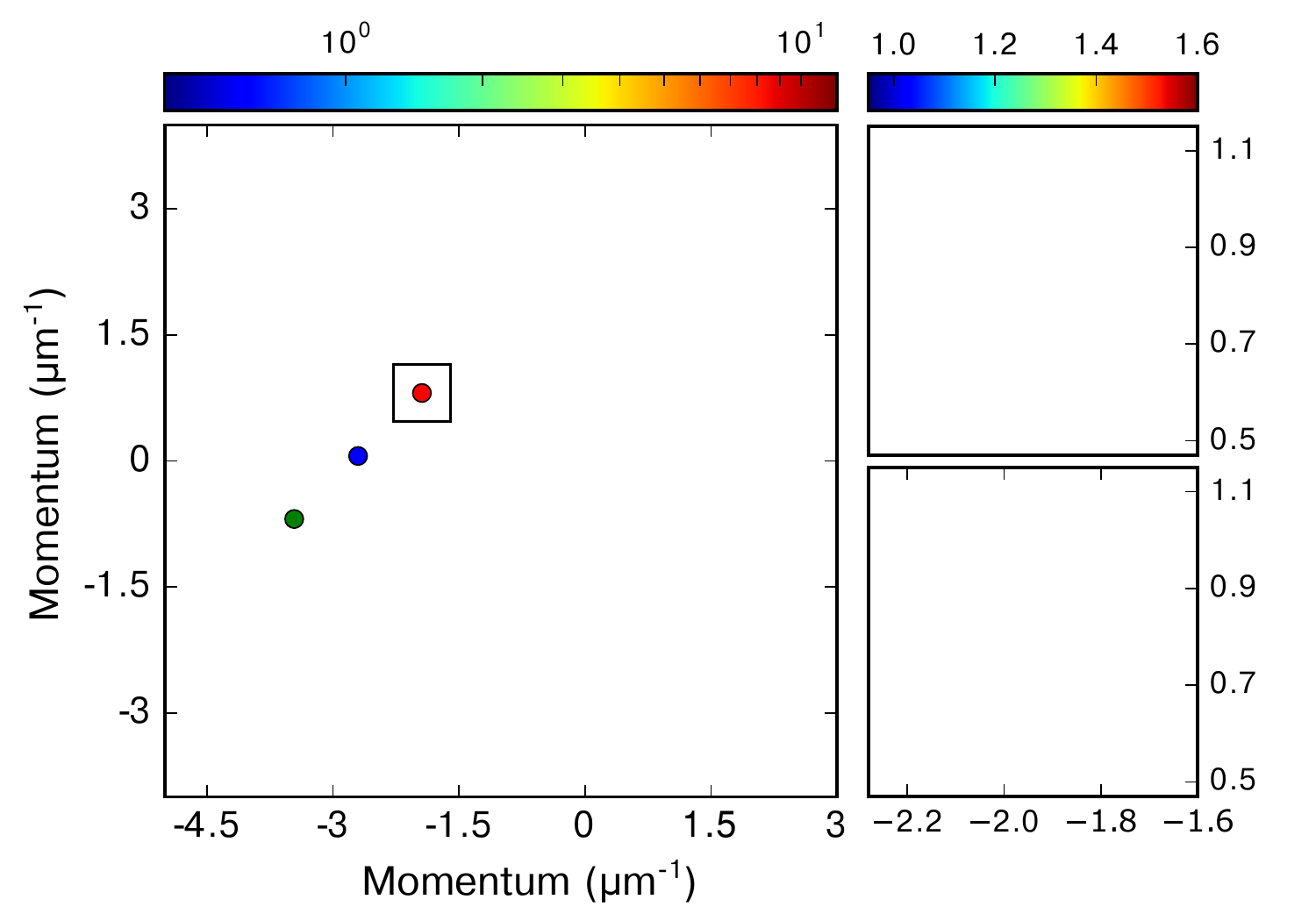}

\caption{Left: Measured logarithmic far-field emission intensity (in arbitrary
units) using \SI{100}{\micro W} pump power with the position of the
pump (blue), signal (red) and idler (green). The shadow is from the
small wrench that blocks the reflected pump light in the back-focal
plane of $\mathrm{L_{1}}$. Right: auto-correlations of the signal
beam at the red circle in the left graph (bottom) and cross-correlations
between signal (red) and idler (green) beams (top) as a function of
the pick-up momentum of the signal mode. The scan range is denoted
by the small square around the signal (red) mode. }

\label{fig:apr23_posscan_posscanauto}
\end{figure}

We measured correlations between two points in the far-field, i.e.
at two defined polariton momenta, as shown on the left hand side of
\figref{apr23_posscan_posscanauto}, which was acquired by a CCD camera
inserted in one of the Fourier planes, \emph{FP}, in \figref{experimental_setup}:
two pump polaritons scatter into a lower and a higher momentum state
on a figure-eight shaped pattern given by energy-momentum conservation
on the lower polariton branch, see also the data in Ref. \cite{Langbein2004,Langbein2004a}.
In what follows, we designate the low momentum state as signal (red),
and the high momentum state as idler (green). The pump polariton direction
has been blocked at the back-focal plane of $\textrm{\ensuremath{L_{1}}}$
to avoid overexposure. Typical focal spot sizes, measured in the real-space
plane between \emph{$\textrm{\ensuremath{L_{2}}}$} and $\mathcal{\textrm{\ensuremath{L_{3}}}}$
are about \SI{80}{\micro m} (including the pinhole diameter), which
leads to a mode size of about \SI{0.1}{\micro m^{-1}}, and thus,
the image in \figref{apr23_posscan_posscanauto} contains approximately
5000 (70 by 70) modes. As mentioned earlier, the fibers are of diameter
\SI{62}{\micro m}. Taking the magnification of our setup into account,
the fibre diameter corresponds to approximately half of a spatial
mode size in the far-field. 

\section{Results\label{sec:results}}

\subsection{General results and first observations\label{sub:General-results-and}}

\begin{figure}[h]
\centering%
\begin{minipage}[c]{0.95\columnwidth}%
\includegraphics[width=0.98\columnwidth]{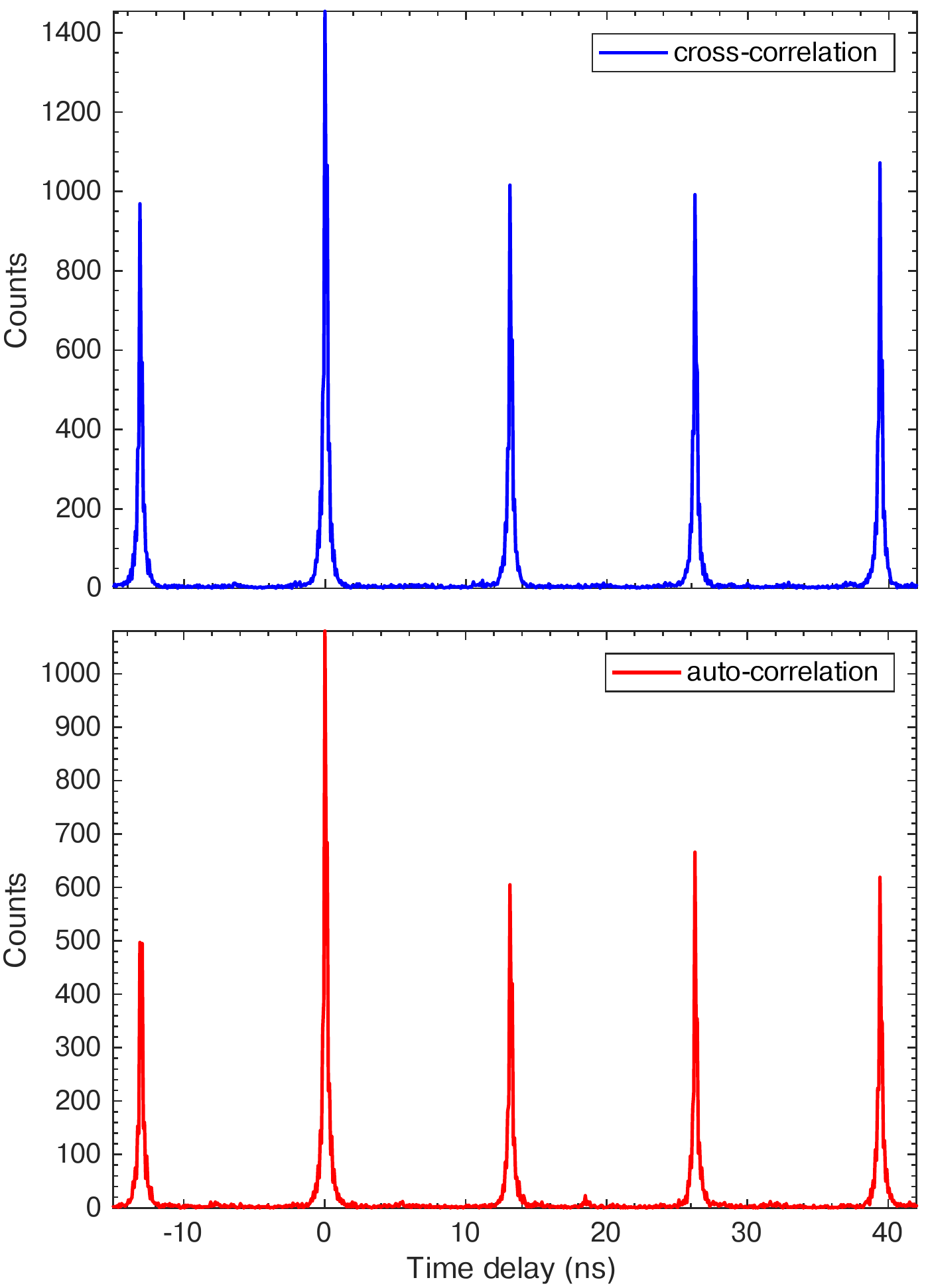}%
\end{minipage}

\caption{Typical correlation histograms, where the signal mode was set at $k_{\mathrm{S}}=\left(-2.34,1.6\right)\SI{}{\per\micro\metre}$
and idler mode at $k_{\mathrm{S}}=\left(-3.5,-0.84\right)\SI{}{\per\micro\metre}$.
The cross-correlation is shown at the top and auto-correlation of
the signal mode at the bottom. The excitation wavelength is 815.6\,nm
and the pump power is \SI{32}{\micro W}.}

\label{fig:feb16_power_0045_0046}
\end{figure}

A typical correlation histogram, between the signal-signal and signal-idler
modes is shown in \figref{feb16_power_0045_0046}. The histograms
contain a train of pulses of nearly constant height, except for the
one at zero delay, which is a sign of correlation. The pulses are
separated by the repetition rate of the excitation laser. As a measure
of correlations, we use the ratio of the area of the peak at zero
time relative to the area of the peaks at $\pm13$\,ns, which corresponds
to the second-order coherence (\eqref{gsidef}) in the limit of low
efficiencies \cite{Stevens:14} (we measured total efficiencies of
roughly $0.4$\, \% for all channels). The peak areas are computed
by integration over a time window as wide as the repetition period
(in this case $13\,$ ns). A value of one corresponds to coherent
light, and it is evident from the figure that we observe bunching
at zero delay. 

The cross-correlations above do not exist for modes that do not fulfill
the phase matching condition. For the measurement on the right hand
side of \figref{apr23_posscan_posscanauto}, we kept the momentum
of the idler mode fixed and scanned with the signal fibre over a momentum
range of about $0.7\times$\SI{0.7}{\micro m^{-2}}, as indicated
by the small black square in the far-field image (\ref{fig:apr23_posscan_posscanauto},
left). At each fibre position, we plot the value of the correlations.
The cross-correlation on the top right hand side of \figref{apr23_posscan_posscanauto}
(signal-idler) reduces to one at about \SI{0.2}{\micro m^{-1}} away
from the phase matching in any direction, which is equivalent to a
Fourier plane shift of four times the fibre diameter. Given that both
the mode size and the fiber diameter correspond to about \SI{0.1}{\micro m^{-1}},
a simple convolution would lead to an apparent momentum range of about
\SI{0.15}{\micro m^{-1}}, slightly smaller than what we measure.
We note that similar results were obtained by reversing the roles,
fixing the signal momentum, and scanning with the idler, and also
when we measured the auto-correlation (signal-signal) by setting both
fibers to either the signal momenta, and scanning with one of them,
as shown on the right hand side bottom panel of \figref{apr23_posscan_posscanauto}.
In the case of auto-correlations we obtain a measurement of the mode
size in k-space, whereas the cross-correlation measurement is sensitive
to both the phase-matching condition and the mode size.

\begin{figure}[h]
\includegraphics[width=0.98\columnwidth]{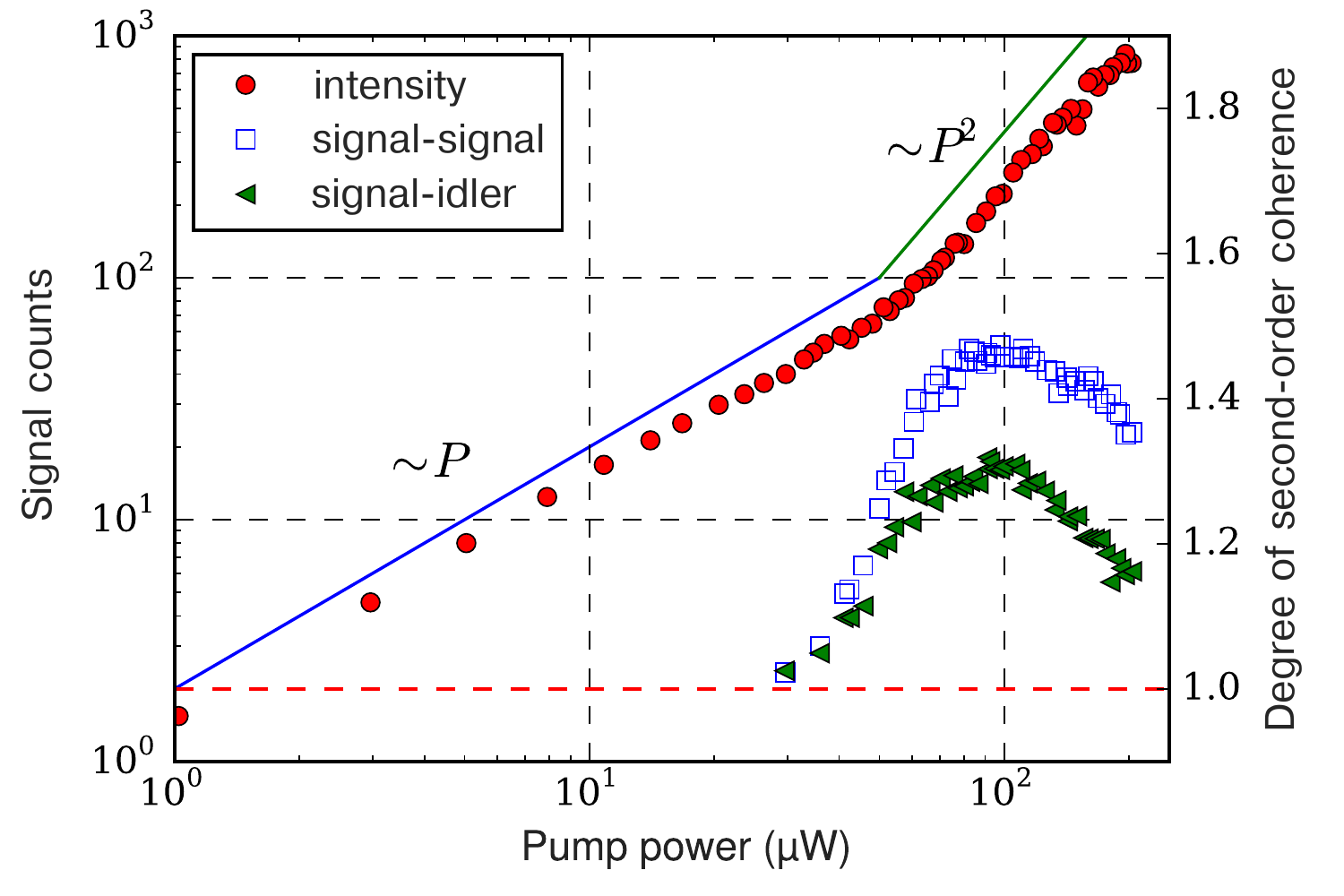}

\caption{Luminescence intensity of the signal of \figref{apr23_posscan_posscanauto}
(left axis, solid red circles), and signal-signal/signal-idler correlations
(right axis, open blue squares/solid green triangles) as a function
of the excitation power. The excitation wavelength is 815.8\,nm.
Uncertainties for the correlations, are in the 0.02-0.05 range, while
for the intensity, they are smaller than the symbols. }

\label{fig:feb16_power_CCD_bunching}
\end{figure}

It is also clear from \figref{feb16_power_0045_0046} that the auto-correlation
is higher than the cross-correlation. The figure does not display
the idler-idler histogram, which is very similar to the signal-signal
histogram, but with lower count rates, because the idler mode is more
excitonic and therefore, dimmer. It turns out that for almost all
excitation powers the auto-correlation is larger than the cross-correlation,
as demonstrated in \figref{feb16_power_CCD_bunching}. Only at low
pump powers the correlations are similar in magnitude. Similar results
were reported for 1D microcavities by Ardizzone et al. \cite{Abbarchi2012}.
Also shown in \figref{feb16_power_CCD_bunching} is a the corresponding
emission intensity at the signal momentum as a function of excitation
power. First, the intensity increases linearly, and at a certain threshold
power it becomes quadratic, indicating that polaritons responsible
for the emission are generated by the interaction Hamiltonian in \eqref{four-wave-mixing-Hamiltonian}.
The intensity at the idler position (not shown) behaves in the same
way with the same threshold, except that its value is reduced due
to the lower photon content. 

Figure \ref{fig:feb16_power_CCD_bunching} also demonstrates that
the correlations become larger than one only, when the pair intensity
starts to overcome the uncorrelated background. We should mention
here that the maximum of the correlations, which strongly depends
on the exact location on the sample, always occurs at a somewhat higher
excitation power than the threshold power between the linear and quadratic
dependencies. As seen in \figref{feb16_power_CCD_bunching}, for this
particular position, the threshold position is at around \SI{60}{\micro W},
while the correlations reach their maximum at around \SI{90}{\micro W}.
A similar behavior was found in \cite{Ardizzone2012,Ardizzone2012a,Ardizzone2013a}.
What is also clear from this plot, is that the bunching has a rather
sharp threshold, above which it increases nearly linearly till it
reaches a maximum, and then drops slowly. As the pump intensity increases,
the polariton dispersion is renormalized because of the density-dependent
repulsive interactions between polaritons \cite{Ciuti2003}. In order
to avoid spurious effects related to this renormalisation, we always
re-align the fibers in such a way that the bunching was highest. The
slow drop in the correlations can be attributed to the presence of
more than one polariton in the cavity and the creation of multiple
pairs \cite{McNeil1983}. Similar conclusions were drawn in \cite{Ardizzone2012}.

\begin{figure}[h]
\includegraphics[width=0.98\columnwidth]{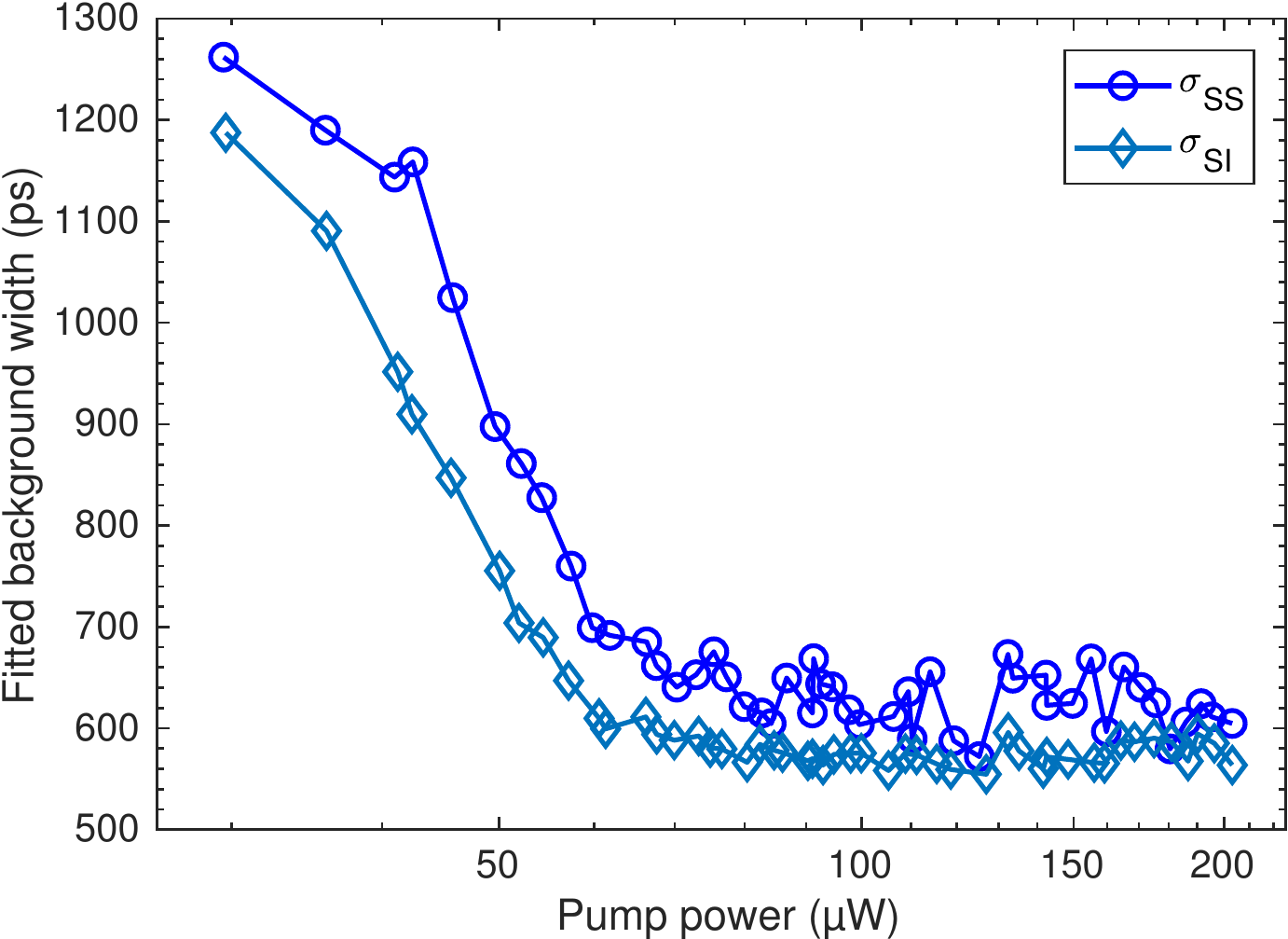}

\caption{Temporal width of the background extracted when fitting the central
peak of the histograms as in \figref{feb16_power_0045_0046} with
double-Gaussian distributions, for both the auto- and cross-correlations
(as labeled). The underlying data is the same as for \figref{feb16_power_CCD_bunching}.
For the fit we assume that the parametrically scattered polaritons
have a constant width given by two-times the time resolution of the
detection system, in this case we have $\sigma_{\mathrm{pol}}\equiv2\tau=324\,$ps.}

\label{fig:backgroundwidth}
\end{figure}

The low pump power regime is dominated by background from bound state
excitons \cite{Langbein2004a} that trap the initially generated free
polaritons until all impurities in the excitation region are saturated,
which occurs approximately at \SI{50}{\micro W} pump power as can
be seen in \figref{backgroundwidth}. Unfortunately the energy and
momentum range of these bound states has a large overlap with the
experimentally accessible range for signal and idler polaritons \cite{Langbein2004a}
and it is therefore not possible to simply filter out all the background
light. We, however, expect the signal-to-background ratio to improve
if we additionally apply filtering in the time domain, because the
radiative lifetime of the bound states is more than a factor of $5$
longer than the polariton lifetime \cite{Langbein2004a}. We analyze
two and three photon measurements using the highest time resolution
we could achieve in more detail in the following section.

\subsection{High time resolution coincidence measurements\label{sub:High-time-resolution}}

From the results in the previous section we can deduce that the optimal
pump power lies somewhere below $80$\,$\mu$W and possibly even
significantly lower when reducing the background further through temporal
filtering. The goal is to find a setting where we can reduce the power
as much as possible while still retaining some parametric signal above
the background. We then record coincidence events with the HBT setup
described above, but replace the APD with superconducing nanowire
single photon detectors (SNSPDs) in order to achieve a higher total
time resolution of $\approx35\,$ps per channel. For the data presented
in this section we use a bin width of $\SI{4}{ps}$ and again a pulsed
laser as pump.

In the following analysis we use, unless otherwise specified, a dataset
obtained with the following settings. We have set a pump power of
\SI{32}{\micro W} at $1.5204\,$eV and excite with $k_{\mathrm{p}}=$\SI{1.34}{\micro m^{-1}}.
Through fitting a three oscillator model \cite{Langbein2004} we find
the photon-exciton detuning to be $\Delta_{\mathrm{C-X}}(0)=-2.2\,$meV
and the heavy hole exciton energy is $E_{\mathrm{HH}}=1.52169\,$eV.

\begin{figure}[h]
\centering%
\begin{minipage}[t]{0.95\columnwidth}%
\begin{center}
\includegraphics[width=0.98\columnwidth]{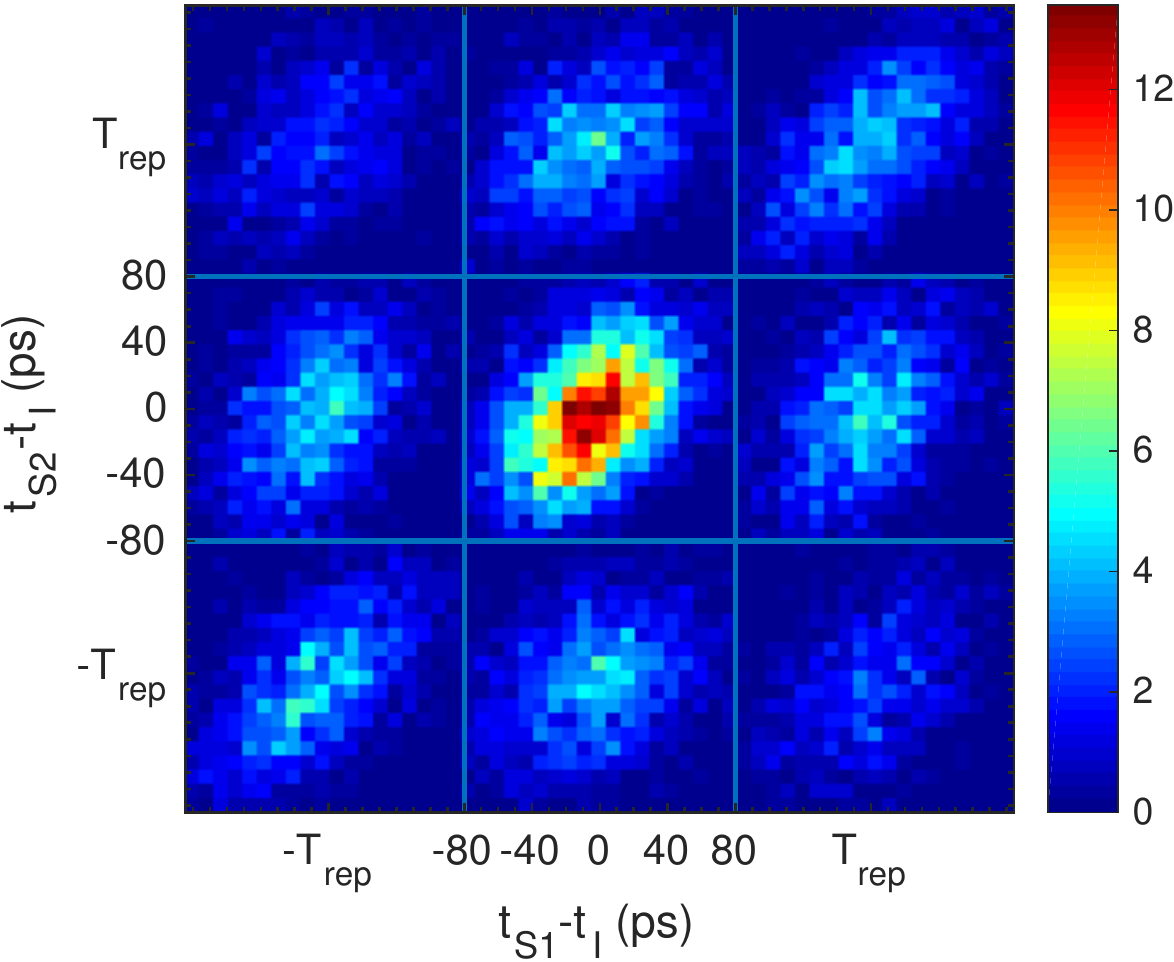}
\par\end{center}%
\end{minipage}\caption{Triple coincidence histogram in a $3\times3$-grid of nearest neighbor
pulses around $t_{\mathrm{S1}}=t_{\mathrm{S2}}=t_{\mathrm{I}}$. Each
segment of the grid is subdivided into a 20 by 20 grid consisting
of $\SI{4}{ps}$ bins.}
\label{fig:aug15_fulltriplesgrid}
\end{figure}

Shown in \figref{aug15_fulltriplesgrid} are the recorded triple coincidences
as a function of the relative times between each of the two signal
modes and the idler mode. Plotted are, on a $3\times3$-grid, only
the central $\pm\SI{80}{ps}$ around the pulse maxima (each cell in
the blue grid) for integer multiples of $T_{\mathrm{rep}}=\SI{13.1}{ns}$
(cell spacing), corresponding to the pulse repetition period of the
pump laser. Because of the short lifetime of the polariton states
($T_{\mathrm{c}}<\SI{10}{ps}$) the different regions in the $3\times3$-grid
satisfy the condition $T_{\mathrm{rep}}\gg T_{\mathrm{c}}$ well.

\begin{figure*}[t]
\centering%
\begin{minipage}[t]{0.99\textwidth}%
\begin{center}
\includegraphics[width=0.98\columnwidth]{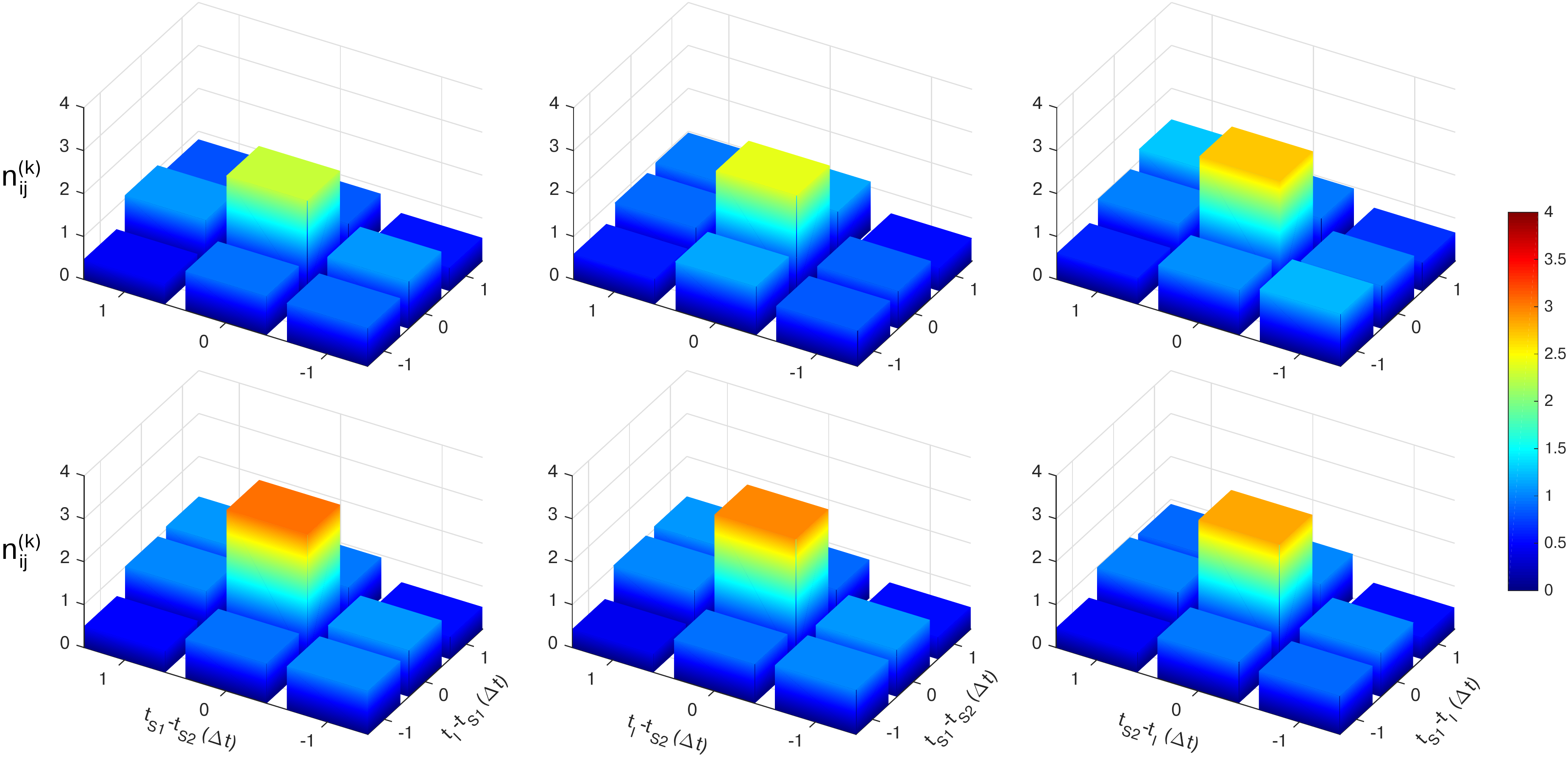}
\par\end{center}%
\end{minipage}

\caption{Measured $n_{ij}^{(k)}$ for two different datasets (rows) for $\tau_{\mathrm{W}}=\SI{12}{ps}$
and binned into a single bin per pulse. From left to right: time delays
$t_{S1}$,$t_{\mathrm{S2}},$ and $t_{\mathrm{I}}$ is kept at zero
delay, respectively, while the other two channels are delayed. Note
that the normalization is done by averaging over the nearest neighbor
of the off-diagonal peaks. The bottom row shows the same dataset as
\figref{aug15_fulltriplesgrid} whereas the top row shows a similar
dataset, at a different sample position, with more background contribution
at a pump power of \SI{37}{\micro W}. All color bars span the same
linear range from $0$ to $4$. }
\label{fig:aug_triples_binned_directions_noalpha}
\end{figure*}
\begin{figure}
\centering%
\begin{minipage}[t]{0.95\columnwidth}%
\begin{center}
\includegraphics[width=0.98\columnwidth]{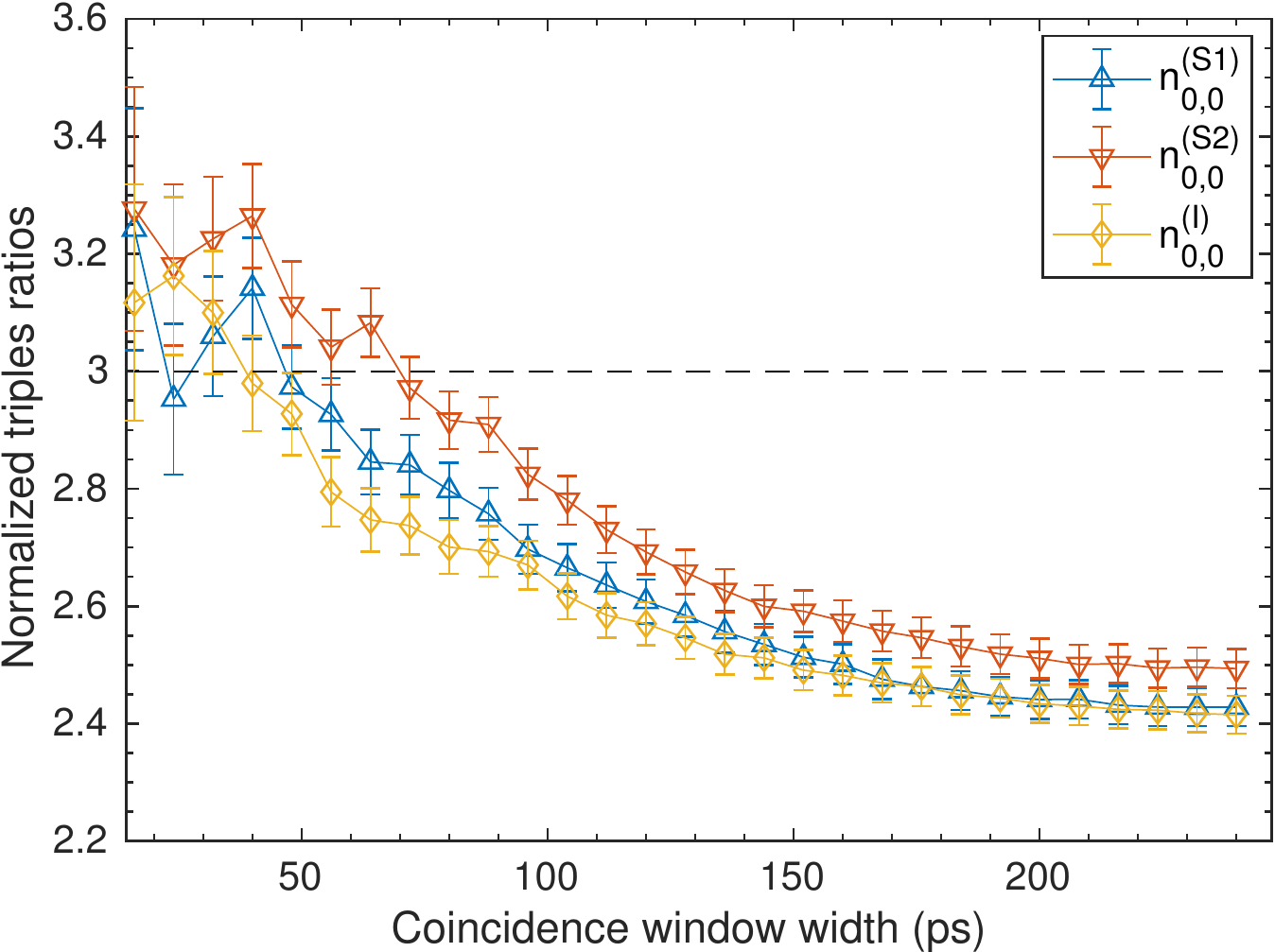}
\par\end{center}%
\end{minipage}\caption{Coincidence window dependence of the normalized triples $n_{0}$ (\eqref{triprat0quant})
for the three different channel permutations. The dashed line indicates
the maximum value reachable using the (classical) high pump limit
(\eqref{triprat0class}).}
\label{fig:aug15_triples_vs_coincwindow}
\end{figure}

Since we know that the background states have lifetimes $T_{\mathrm{th}}\approx\SI{100}{ps}$,
much longer than the polariton states, we filter coincidences around
the center of each pulse with varying coincidence windows $\tau_{\mathrm{W}}$
in order to reduce the background contribution. To capture the influence
of the different channels we additionally analyze the measured coincidences
for all three permutations of channels assignments, i.e.: 
\begin{align}
n_{ij}^{(\mathrm{S1})}\equiv & \:\frac{N_{\mathrm{SSI}}(0,iT_{\mathrm{rep}},jT_{\mathrm{rep}})}{N_{\mathrm{SSI}}(0,T_{\mathrm{rep}},0)},\nonumber \\
n_{ij}^{(\mathrm{S2})}\equiv & \frac{N_{\mathrm{SSI}}(iT_{\mathrm{rep}},0,jT_{\mathrm{rep}})}{N_{\mathrm{SSI}}(0,T_{\mathrm{rep}},0)},\quad n_{ij}^{(\mathrm{I})}\equiv\:n_{ij},
\end{align}
producing delays along different physical channels. Shown in \figref{aug15_triples_vs_coincwindow}
is the dependence of $n_{ij}^{(k)}$ on the coincidence window width
$\tau_{\mathrm{w}}$. In \figref{aug_triples_binned_directions_noalpha}
we compare coarse grained correlation histograms from the dataset
used for \figref{aug15_fulltriplesgrid} with a different dataset
with a higher background contribution. 
Because we permute all channels the overall symmetry changes the ``ridge'',
namely where $t_{\mathrm{S1}}=t_{\mathrm{S2}}$, (see \figref{hsitvspumpparam})
from diagonal (like in \figref{aug15_fulltriplesgrid}) to along one
of the axes. The first dataset (top row) shows the expected classical
signature quite clearly, while the second one shows a more pronounced
central peak with the components on the $t_{\mathrm{S1}}=t_{\mathrm{S2}}$-diagonal
reaching below one, which approaches the non-classical distribution
in \figref{hsitvspumpparam}. We expect the central peak to have the
same height for all permutations, which is also observed apart from
small variations.
\begin{figure*}[t]
\centering%
\begin{minipage}[t]{1\textwidth}%
\begin{center}
\includegraphics[width=1\columnwidth]{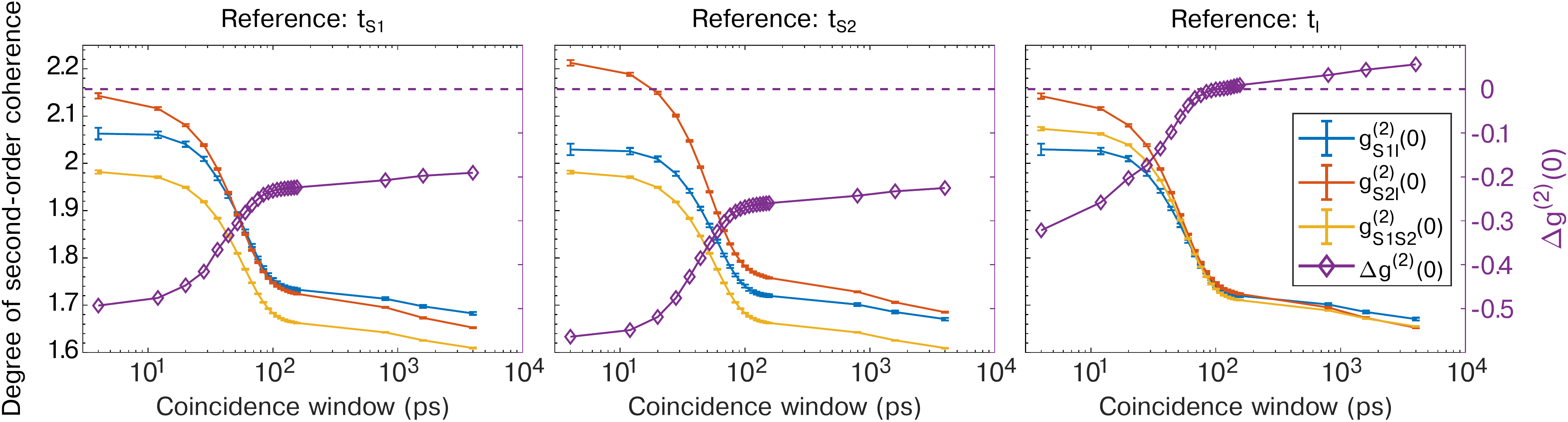}
\par\end{center}%
\end{minipage}\caption{Coincidence window dependence of all three two-detector correlations
for different reference channels, as labeled. From left to right:
$t_{\mathrm{S1}}$,$t_{\mathrm{S2}}$ and $t_{\mathrm{I}}$ is set
as reference, respectively. The right-hand axis (purple) shows the
corresponding $\Delta=\left(g_{\mathrm{S1S2}}^{(2)}(0)\right)^{2}-g_{\mathrm{S1I}}^{(2)}(0)g_{\mathrm{S2I}}^{(2)}(0)$,
which below zero (dashed lines), violates the Cauchy-Schwarz inequality.}
\label{fig:doublescoincwindowdep}
\end{figure*}

Since the height of the central peak is theoretically unbounded in
the case of the parametric amplifier source, while it remains constant
for a classical source, we expect it to rise with decreasing $\tau_{w}$
as soon as enough of the long-lifetime thermal light is removed. We
can see this rise in triples quite clearly as well as it starting
to settle below $\approx\SI{35}{ps}$, which corresponds to the time
resolution of our setup. In addition, we observe that for the lowest
filter values, $n_{0}$ reaches values above three indicating a departure
from the classical parametric regime and the onset of the quantum
regime, as discussed in section \ref{sec:theory}.

In order to gain further insight we can additionally evaluate the
data from coincidences between the combination of only two detectors
(of the same measurement) in a similar way (as described in \subref{General-results-and})
to obtain the degree of second-order coherence along the different
directions: 
\[
g_{\mathrm{AB}}^{(2)}(0)=\frac{N_{\mathrm{AB}}(0)}{N_{\mathrm{AB}}(T_{\mathrm{rep}})}.
\]
The denominator in this case is given by the coincidence histogram
delayed by (multiples of) $T_{\mathrm{rep}}$. This gives either the
second-order auto-correlation function $g_{\mathrm{SS}}^{(2)}$, if
the double coincidences are between the two ``signal''-channels,
or cross-correlation functions $g_{\mathrm{SI}}^{(2)}$ otherwise.
We have not measured the corresponding idler auto-correlation $g_{\mathrm{II}}^{(2)}$
for this particular dataset, but we found from several other measurements
that it takes on values similar to $g_{\mathrm{SS}}^{(2)}$. The dependence
of these quantities on $\tau_{\mathrm{W}}$ is shown in \figref{doublescoincwindowdep}.
We can again see a clear dependence on the coincidence window and
the cross-correlations reaching values $>2$. Assuming that $g_{\mathrm{II}}^{(2)}\approx g_{\mathrm{SS}}^{(2)}$,
which we found in an independent measurement, we estimate the Cauchy-inequality
(\eqref{cauchy}) and find it to be violated for all cases at low
coincidence window widths. 

\begin{figure}[h]
\centering%
\begin{minipage}[t]{0.98\columnwidth}%
\begin{center}
\includegraphics[width=0.98\columnwidth]{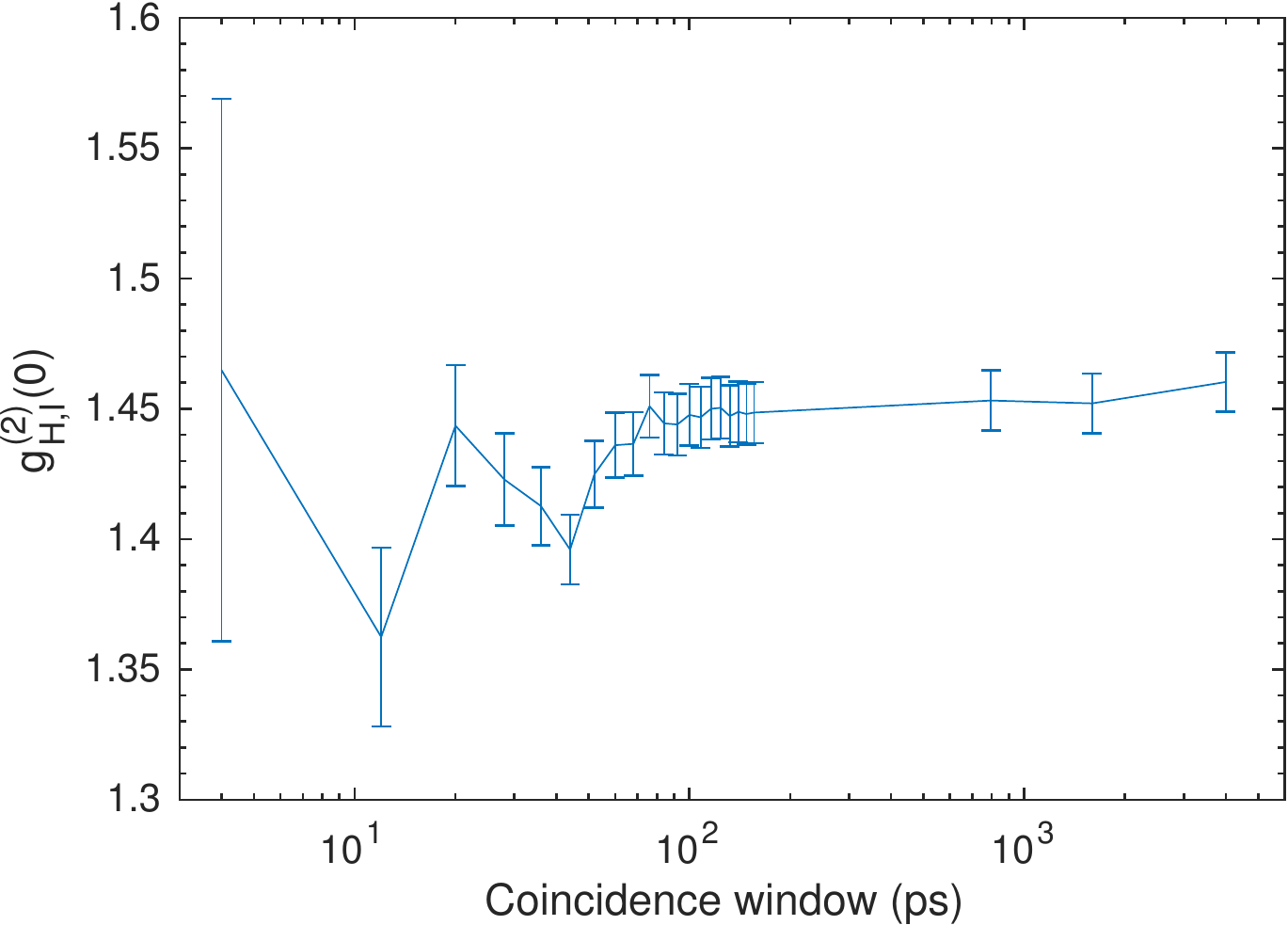}
\par\end{center}%
\end{minipage}\\

\begin{minipage}[t]{0.98\columnwidth}%
\begin{center}
\includegraphics[width=0.98\columnwidth]{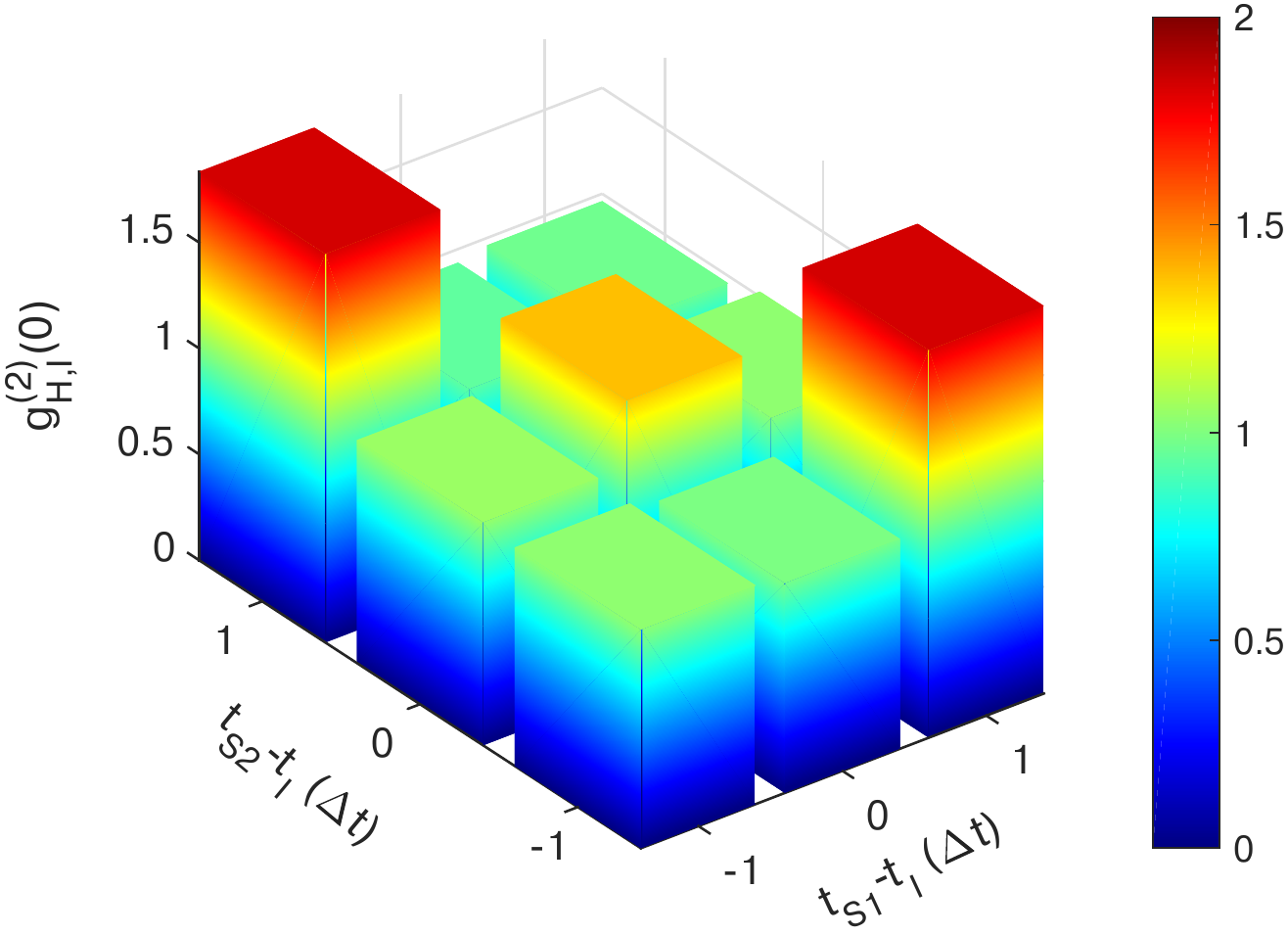}
\par\end{center}%
\end{minipage}\caption{Measured conditional coherence $g_{\mathrm{H}}(0)$ as a function
of the coincidence window width (top figure) and the corresponding
coherences at the lowest value with 12\,ps window width (bottom figure). }
\label{fig:g2hcomp}
\end{figure}

From the cross-correlation value we can also estimate the value of
the heralded coherence function at zero delays (see \cite{0953-4075-42-11-114013}):

\begin{equation}
g_{\mathrm{H}}(0)=\frac{2}{g_{\mathrm{SI}}^{(2)}}\left(2-\frac{1}{g_{\mathrm{SI}}^{(2)}}\right),\label{eq:gcgsi}
\end{equation}
where we take the geometric mean of the two cross-correlation combinations
$g_{\mathrm{SI}}^{(2)}$, often referred to as coincidence-to-accidentals
ratio (CAR), as input, see \figref{g2hcomp}. However, due to the
still weak cross-correlations, the heralded coherence does not reach
values below one, which is another requirement for a quantum source.
In order to reach a truly non-classical value a CAR greater than $2+\sqrt{2}\approx3.4$
would be required. 

Another interesting observation is that the auto-correlations reach
values close to two, which means that the underlying effective mode
number collected by the multi-mode fibers is close to one \cite{Christ2011}
and the background processes have to be emitting into the same (spatial)
mode with thermal statistics. 

We can now combine both double and triple coincidences to compute
the heralded coherence (\eqref{g2hdef}), again by computing the probability
for multi-photon clicks by taking the central peak in \figref{aug_triples_binned_directions_noalpha}
and normalizing it to the off-diagonal ($N_{\mathrm{XXX}}^{0}$) peaks,
i.e.:

\begin{equation}
g_{H,I}^{(2)}\cong\frac{\frac{N_{\mathrm{S1S2I}}(0)}{N_{\mathrm{S1S2I}}^{0}}}{\frac{N_{\mathrm{S1I}}(0)}{N_{\mathrm{S1I}}^{0}}\frac{N_{\mathrm{S2I}}(0)}{N_{\mathrm{S2I}}^{0}}},\label{eq:g2hexp}
\end{equation}
where we took the idler channel as the reference and restrict ourselves
to that case in the following discussion.

The measured coincidence window dependence of $g_{H,I}^{(2)}$ is
shown in \figref{g2hcomp}. Interestingly, the conditional coherence
does not exhibit the pronounced dependence on window size we found
when looking at the coincidences alone because both the three-fold
and two-fold coincidences change in the same way and the effect partially
cancels. The temporal distribution of the more non-classical dataset
can be seen in \figref{g2hcomp} (bottom).
We can see a clear anti-bunching dip in the central peak when going
along the diagonal values, however, the value of the central peak
does not drop below one. The minimal value for the central peak in
the case of the parametric amplifier in the classical regime is $1.5$.
In our measurements we reach values as low as $\approx1.36(3)$ -
more than four standard deviation below $1.5$. This is consistent
with the bunching observed in the cross-correlations, where we find
with \eqref{gcgsi} $g_{SI}^{(2)}\approx2.31$. Even though there
is clear anti-bunching, we want to point out that the obtained values
correspond to super-Poissonian statistics and are still far above
the non-classical limit.

\begin{figure}[h]
\centering%
\begin{minipage}[t]{0.98\columnwidth}%
\begin{center}
\includegraphics[width=0.98\columnwidth]{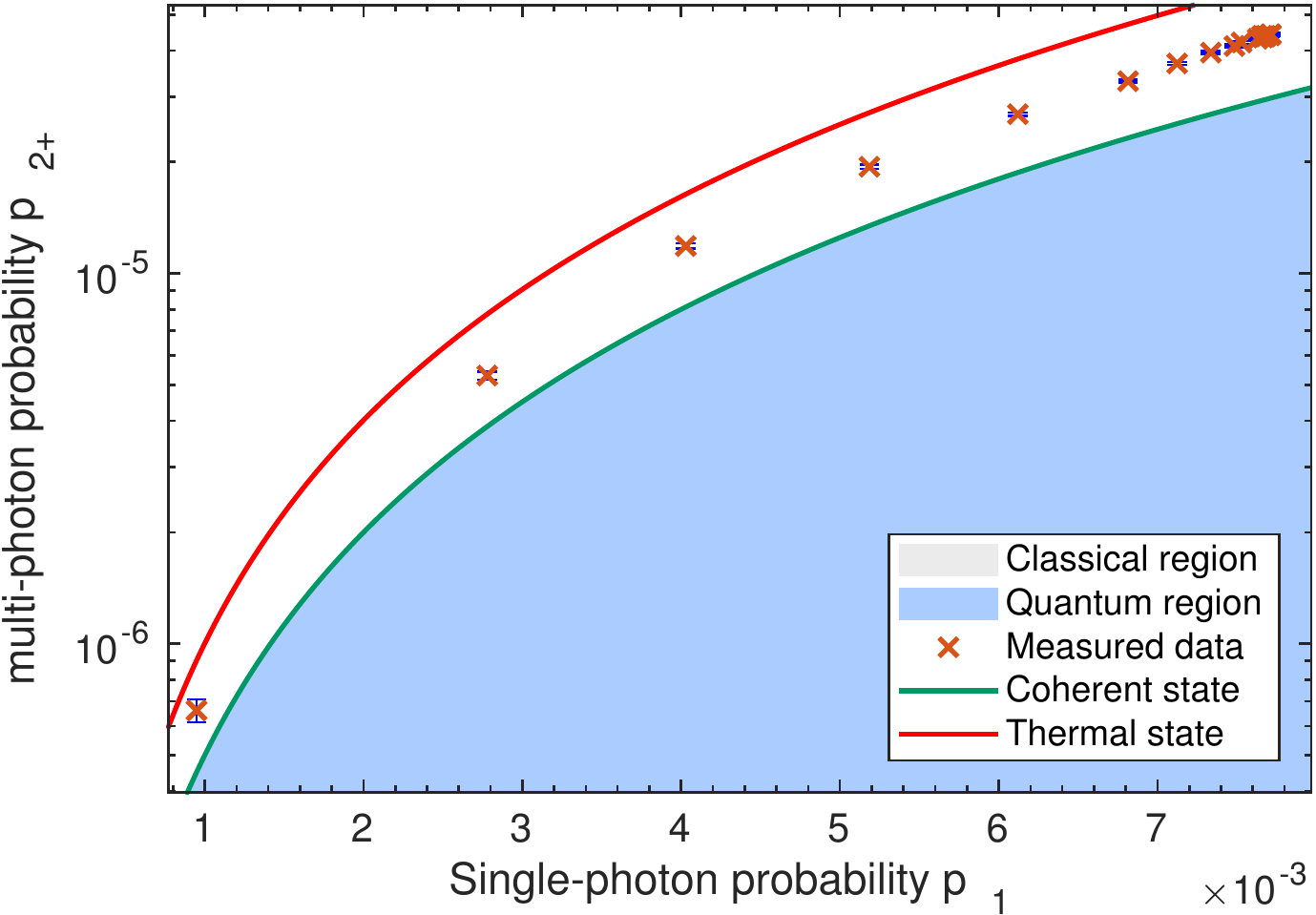}
\par\end{center}%
\end{minipage}\\
\begin{minipage}[t]{0.98\columnwidth}%
\begin{center}
\includegraphics[width=0.98\columnwidth]{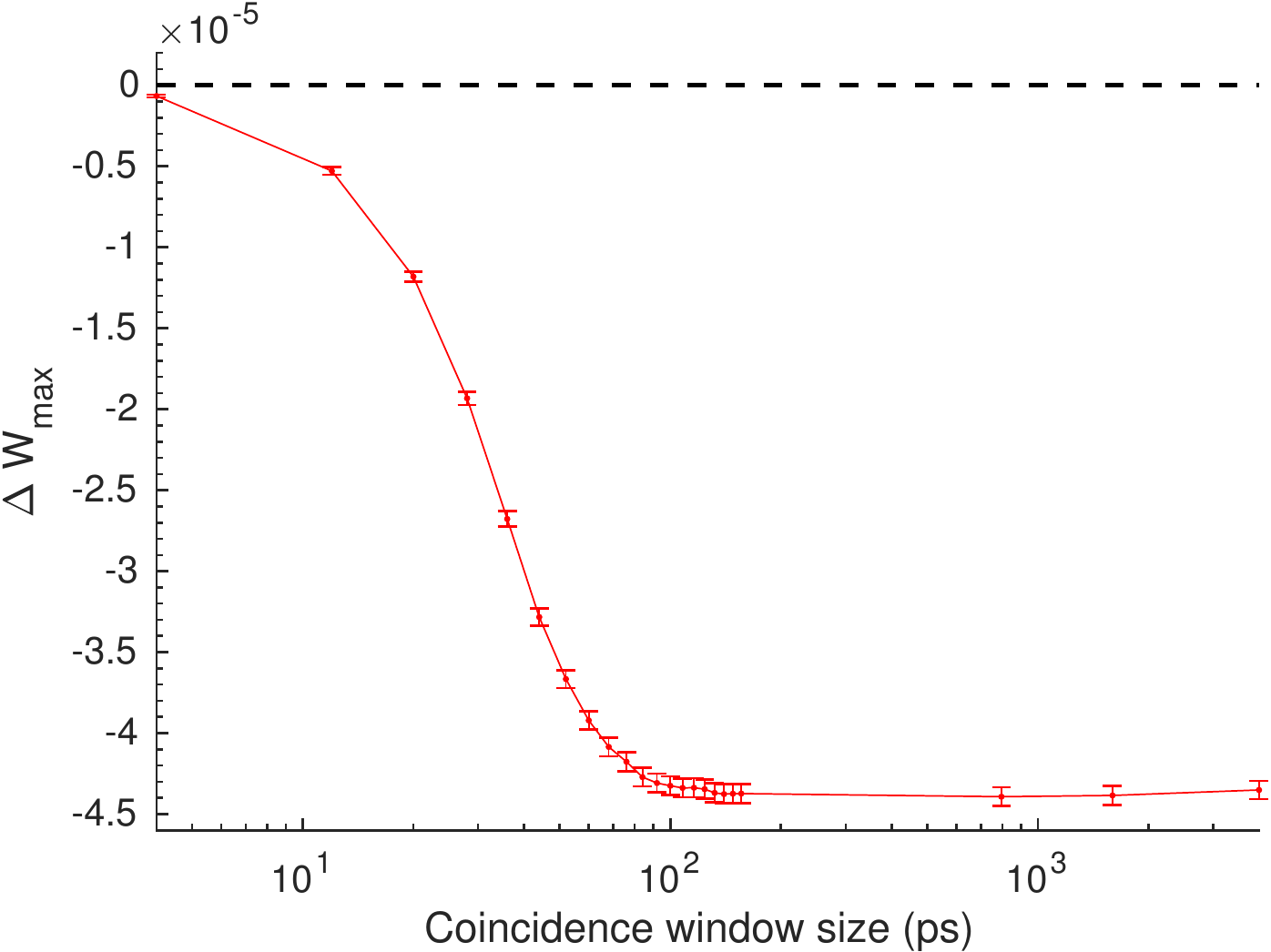}
\par\end{center}%
\end{minipage}\caption{Multiphoton probabilities as a function of single photon probabilities
for the same dataset as in \figref{g2hcomp} plotted in the top figure.
Plotted in the bottom figure is the maximum distance $\Delta W$ (given
by \eqref{wittdistmax}) according to the non-classicality witness
for the same data. A value $\leq0$ signifies that we are in the classical
domain, while positive values signify the quantum domain.}
\label{fig:witts}
\end{figure}

For the final analysis we want to use the non-classical character
witness defined in \subref{Non-classicality-characterizatio} and
compare the single photon and multiphoton probabilities of our source.
We compute the probabilities, as defined in \subref{Non-classicality-characterizatio},
as a function of the coincidence window width and plot them in the
same way as in \figref{spmptheory}, resulting in the plot shown in
\figref{witts}.
Interestingly for this measure the coincidence filtering seems to
have a much greater effect than for the conditional coherence as can
be seen in the right-hand figure in \figref{witts}. By comparing
the calculated probabilities to ideal states we can see that the underlying
quantum state always lies somewhere in-between a classical thermal
state and the coherent state and is therefore classical.

While filtering reduces the distance to states with non-classical
character $\Delta W$ (given by \eqref{wittdistmax}) significantly,
we find it to be still approximately $2.6$ standard deviations away
from the boundary to the non-classical region at the smallest coincidence
window, which is roughly what we expect from the value of the conditional
coherence.

\section{Possible improvements and conclusions\label{sec:conclusions}}

The main contribution to the bunching at zero delay comes from the
thermal background, which is difficult to remove. In order to significantly
reduce the overlap with the parametrically scattered polariton states
of the state space, a stronger Rabi splitting with simultaneously
narrower linewidths could help. Based on our structure this would
require an increase of the splitting by at least $1\,$meV ($V_{\mathrm{RS}}=3.5\,$meV
\cite{Jensen2000}), while still retaining the lowest possible free
exciton linewidth. This could be achieved by using two $25\,$nm GaAs
quantum wells in the cavity. 

Another limiting factor in our measurements is the time resolution
of the the detection system, which is still roughly a factor of $3$
greater than the lifetime of the polariton states. If we assume that
all detectors have a Gaussian response function we can extrapolate
the increase in coherence we got between measurements performed with
avalanche photo diodes with $45\,$ps FWHM (\figref{feb16_power_CCD_bunching})
and the SNSPD detectors. We find that a detection system with a total
timing jitter of $10\,$ps FWHM could almost double the measured cross-correlation
value giving rise to non-classical anti-bunching (using \eqref{gcgsi}
we get $g_{\mathrm{H}}(0)\approx0.88$). Reaching such low jitter
values might be possible in the near future \cite{EsmaeilZadeh2018}.
A higher detector time resolution would at the same time also help
separating the signal from the background in time, which may yield
a small additional increase in cross-correlations. 

A different approach would be to confine polaritons in all three dimensions
and use interference effects in coupled photonic dots to create entanglement
\cite{Liew2013a}. Quantum confinement additionally gives another
handle to control the energy of the polariton states and at the same
time, slightly increases the interaction strength. Very recently the
successful generation of non-classical polariton states in a single
photonic dot was reported \cite{Munoz-Matutano2017,Delteil2018}.

In conclusion, we have demonstrated that correlated photon pairs can
be generated by means of polariton parametric scattering in planar
microcavities. In particular, we have shown that the signal-signal,
and idler-idler correlations are consistently higher than the signal-idler
correlations. This finding is in stark contrast with the predictions
of models based on simple four-wave mixing processes. We can explain
the differences to the theoretical models by including uncorrelated
background light. We attribute this background light to donors resonantly
absorbing and re-emitting free excitons (polaritons) in the same phase-space
region as the modes of the polariton parametric amplifier. We measured
that with appropriate spatiotemporal selection at very low pump powers
the photon statistics changes significantly. Fluctuations in photon
numbers increase in both the triples and doubles measurements, anti-bunching
gets more pronounced and the distance to quantum non-Gaussian emission
characteristic is decreased significantly, but unfortunately, the
values still stay in the classical regime. We consider different improvements
that could reduce the background further and find that an increase
in time resolution of the detection system would suffice to observe
non-classical photon statistics. 
\begin{acknowledgments}
We would like to thank I. Carusotto, D. Pagel, A. Imamo{\u g}lu for
stimulating discussions. We acknowledge financial support by the European
Union Research Council (ERC) EnSeNa, grant number 257531, and the
Austrian Science Fund (FWF), grant numbers P-22979-N16 and I2199-N27,
and the UK EPSRC grant EP/M013472/1. Appendix A provides the experimental
data for the central results of this paper. 
\end{acknowledgments}

\bibliographystyle{apsrev4-1}
\bibliography{B:/Data/Polariton/Papers/polariton/polariton,polariton}

\begin{widetext}\appendix

\section*{Appendix A}

Table 1 provides the raw data extracted directly from a 15 hours time-tag
measurement. This data can be used to reproduce the traces in figures
15 and 16,  and we used $N_{0}=\frac{N_{\mathrm{S1S2I}}^{0}}{N_{\mathrm{S1I}}^{0}N_{\mathrm{S2I}}^{0}}$: 

\vspace{5mm}

\begin{table}[h]
\centering%
\begin{tabular}{|c|c|c|c|c|c|c|c|}
\hline 
$\tau_{\mathbf{W}}$ & $N_{0}$ & $N_{\mathrm{S1I}}$ & $N_{\mathrm{S2I}}$ & $N_{\mathrm{S1S2I}}$ & $\Delta W$ & $\delta\left(\Delta W\right)$ & $g_{\mathrm{H}}(0)$\tabularnewline
\hline 
4 & 380986080 & 304359 & 57152 & 67 & -2.1e-7 & 8.1e-8 & 1.465\tabularnewline
\hline 
12 & 379723082 & 888105 & 168648 & 540 & -1.42e-6 & 2.28e-7 & 1.362\tabularnewline
\hline 
20 & 416606617 & 1411497 & 270937 & 1335 & -3.68e-6 & 3.25e-7 & 1.444\tabularnewline
\hline 
28 & 424958538 & 1850823 & 358853 & 2245 & -5.85e-6 & 4.1e-7 & 1.423\tabularnewline
\hline 
36 & 427447870 & 2194373 & 430213 & 3155 & -7.99e-6 & 4.8e-7 & 1.413\tabularnewline
\hline 
44 & 428759188 & 2447406 & 484692 & 3910 & -9.55e-6 & 5.29e-7 & 1.396\tabularnewline
\hline 
52 & 440205249 & 2622812 & 524654 & 4513 & -0.00001121 & 5.5e-7 & 1.425\tabularnewline
\hline 
60 & 446834339 & 2738787 & 551939 & 4925 & -0.00001221 & 5.64e-7 & 1.436\tabularnewline
\hline 
68 & 450183896 & 2812835 & 570090 & 5189 & -0.00001274 & 5.72e-7 & 1.437\tabularnewline
\hline 
76 & 455177765 & 2859698 & 581303 & 5375 & -0.00001332 & 5.75e-7 & 1.451\tabularnewline
\hline 
84 & 453838031 & 2889051 & 588254 & 5487 & -0.00001349 & 5.82e-7 & 1.444\tabularnewline
\hline 
92 & 454860004 & 2908194 & 592454 & 5549 & -0.0000136 & 5.83e-7 & 1.444\tabularnewline
\hline 
100 & 456618971 & 2921304 & 595115 & 5592 & -0.00001373 & 5.83e-7 & 1.448\tabularnewline
\hline 
108 & 457247663 & 2931057 & 596773 & 5615 & -0.00001375 & 5.84e-7 & 1.447\tabularnewline
\hline 
116 & 458948286 & 2938524 & 597919 & 5632 & -0.00001382 & 5.83e-7 & 1.45\tabularnewline
\hline 
124 & 459435903 & 2944550 & 598754 & 5647 & -0.00001385 & 5.83e-7 & 1.45\tabularnewline
\hline 
132 & 458480851 & 2949705 & 599453 & 5663 & -0.00001386 & 5.86e-7 & 1.447\tabularnewline
\hline 
140 & 458979902 & 2953990 & 600035 & 5677 & -0.00001392 & 5.86e-7 & 1.449\tabularnewline
\hline 
148 & 459510005 & 2957740 & 600526 & 5679 & -0.00001389 & 5.86e-7 & 1.448\tabularnewline
\hline 
156 & 460134970 & 2961157 & 600946 & 5684 & -0.0000139 & 5.85e-7 & 1.449\tabularnewline
\hline 
796 & 470277911 & 3032347 & 610362 & 5803 & -0.00001406 & 5.82e-7 & 1.453\tabularnewline
\hline 
1596 & 475778960 & 3067924 & 615553 & 5848 & -0.00001401 & 5.78e-7 & 1.452\tabularnewline
\hline 
3996 & 487707513 & 3126042 & 624531 & 5931 & -0.00001407 & 5.7e-7 & 1.46\tabularnewline
\hline 
\end{tabular}

\caption{Experimental raw data used for computing the conditional coherence
and non-classical character witnesses.}
\end{table}

\appendix

\section*{Appendix B\label{sec:Appendix-B}}

In this paper we show data from two different detector types, avalanche
photodiodes from Micro-Photon Devices (PDM Series) and superconducting
nanowire detectors from Single Quantum. The corresponding time responses
of one of the detectors is shown in \figref{detresponses}. The time
response of the SNSPDs is almost ideally Gaussian and significantly
narrower.

\begin{figure}
\centering%
\begin{minipage}[t]{0.5\columnwidth}%
\begin{center}
\includegraphics[width=0.98\columnwidth]{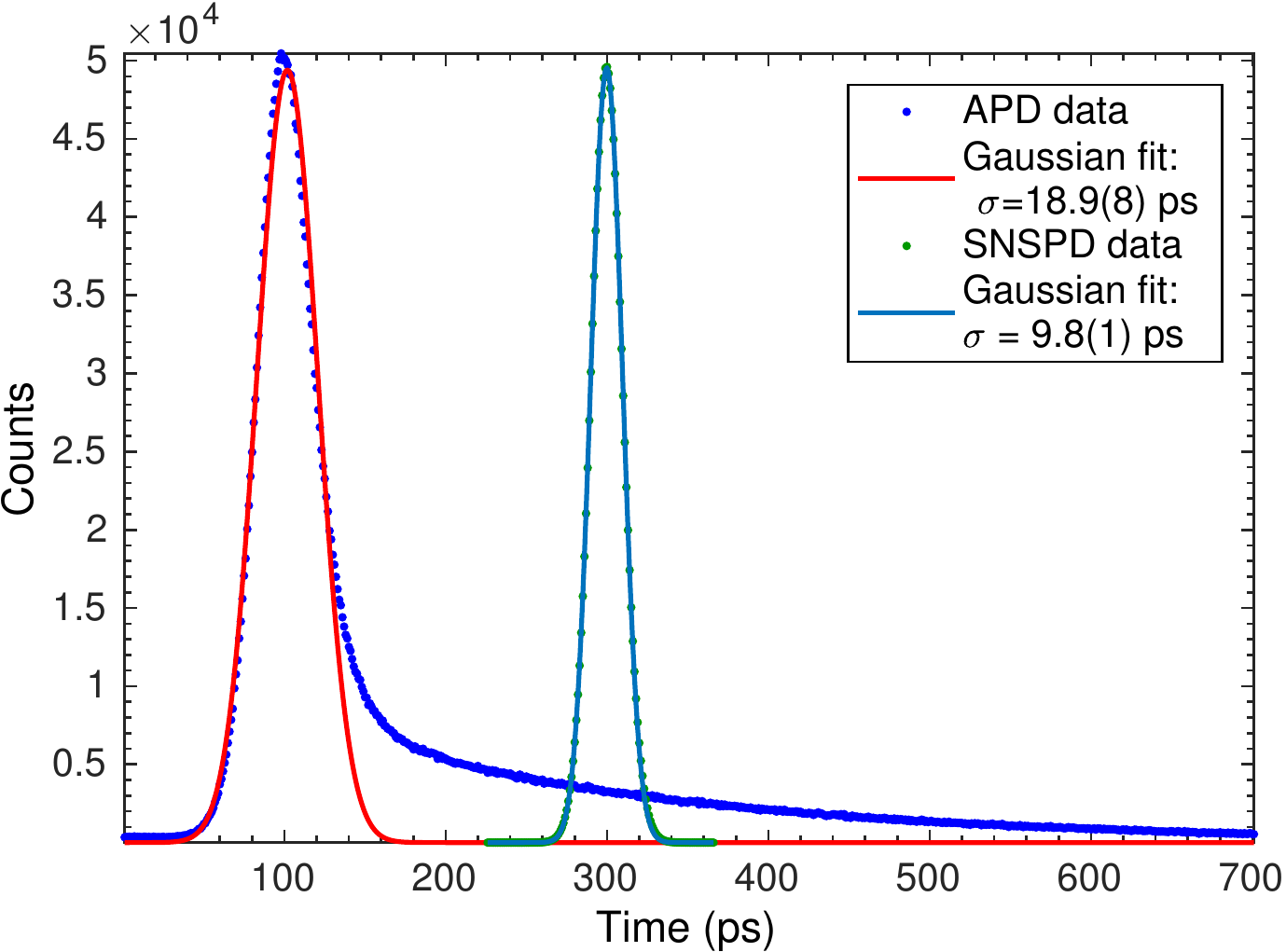}
\par\end{center}%
\end{minipage}\caption{Total time response of two different detectors, as labeled. Both responses
include the time response of the counting system.}
\label{fig:detresponses}
\end{figure}

\end{widetext}
\end{document}